\documentclass[12pt]{article}

\usepackage[a4paper,left=28mm,right=25mm,top=30mm,bottom=30mm]{geometry}

\usepackage{epsfig}
\usepackage{wrapfig}
\usepackage{enumerate}
\usepackage{amsfonts}
\usepackage{amssymb}
\usepackage{latexsym,amssymb}
\usepackage{amsmath}
\usepackage{bm} 
\usepackage{here}

\newcommand{\fracshalf}{\mbox{$\frac{1}{2}$}}

\newcommand{\fracssixteenth}{\mbox{$\frac{1}{16}$}}

\newcommand{\nn}[0]{\hspace*{.7em}}
\newcommand{\n}[0]{\hspace*{.35em}}
\newcommand{\fourl}{\nn \nn}
\newcommand{\T}{^{\mathrm{T}}}

\newcommand{\Gcon}{$G^N_{\mathrm{con}}\,$}
\newcommand{\Gcov}{$G^N_{\mathrm{cov}}\,$}
\newcommand{\Greg}{$G^N_{\mathrm{reg}}\,$}
\newcommand{\Gpar}{$G^N_\pa\,$}

\newcommand\cov{\mathrm{cov}}
\newcommand{\snode}{\mbox {\large 
{$\mbox{$\circ$}$}}}

\newcommand{\ful}{\mbox{$\, \frac{ \nn \nn \;}{ \nn \nn}$}}
\newcommand{\fra}{\mbox{$\hspace{.05em} \frac{\nn \nn}{\nn}\!\!\!\!\! \succ \! \hspace{.25ex}$}}
\newcommand{\fla}{\mbox{$\hspace{.05em} \prec \!\!\!\!\!\frac{\nn \nn}{\nn}$}}
\newcommand{\dal}{\mbox{$  \frac{\n}{\n}\frac{\; \,}{\;}  \frac{\n}{\n}$}}
\newcommand{\fuls}{\scriptsize{\mbox{\raisebox{-.022ex}{$\, \frac{ \nn \n \;}{ \nn \n}$}}}}
\newcommand{\flas}{\scriptsize{\mbox{$\hspace{.05em} \prec\!\!\!\!\!\frac{\nn \nn}{\nn}$}}}
\newcommand{\dals}{\scriptsize{\mbox{$  \frac{\n}{\n}\frac{\; \,}{\;}  \frac{}{}$}}}

\newcommand{\pa}{\mathrm{par}}
\newcommand{\ci}{\mbox{\protect{ $ \perp \hspace{-2.3ex} \perp$ }}}

\newcommand{\txt}{\textstyle}

\newcommand{\E}{{\it E}}
\newcommand{\acal}{\ensuremath{\mathcal{A}}}

\newcommand{\ecal}{\ensuremath{\mathcal{E}}}
\newcommand{\scal}{\ensuremath{\mathcal{S}}}

\newcommand{\pcal}{\ensuremath{\mathcal{P}}}

\newcommand{\mcal}{\ensuremath{\mathcal{M}}}
\newcommand{\ncal}{\ensuremath{\mathcal{N}}}
\newcommand{\hcal}{\ensuremath{\mathcal{H}}}
\newcommand{\ical}{\ensuremath{\mathcal{I}}}
\newcommand{\kcal}{\ensuremath{\mathcal{K}}}

\newcommand{\qcal}{\ensuremath{\mathcal{Q}}}
\newcommand{\wcal}{\ensuremath{\mathcal{W}}}

\newcommand{\vcal}{\ensuremath{\mathcal{V}}}

\newcommand{\sV}{\textsf{V\;}}
\newcommand{\sVs}{\textsf{V}s\;}
\newcommand{\sU}{\textsf{U}\;}
\newcommand{\sUs}{\textsf{U}s\;}

\newcommand{\node}{\mbox {\LARGE
{$\mbox{$\circ$}$}}}
\newcommand{\margnn}{\mbox {\Large
{$\not \: \not $}}$\node $}
\newcommand{\margn}{\mbox {\raisebox{-.1 ex}{\margnn}}}
\newcommand{\condnnc}{\mbox {\LARGE{$\,
\mbox{$\Box$}
\raisebox{.21ex}{\hspace{-1.43ex}\mbox{$\circ$}}
$}} }
\newcommand{\condnc}{\mbox{\raisebox{-.7ex}{\condnnc}}}

\newcommand{\In}{\mathrm{In}}
\newcommand{\inv}{\mathrm{inv}}
\newcommand{\zer}{\mathrm{zer}}

\newtheorem{prop}{Proposition}
\newtheorem{coro}{Corollary}
\begin{document}
\noindent In: (Balakrishnan, N. et al., eds) {\textit{Wiley StatsRef: Statistics Reference Online} (2015); to appear;
also on ArXiv: 1505.02456\\[2mm]

\noindent{{\begin{center}{\large \bf Graphical Markov models, unifying results and their interpretation}\end{center}}
\noindent {\begin{center}{\bf{Nanny Wermuth}} \end{center}

 {\em Mathematical Sciences, Chalmers University of Technology, Gothenburg, Sweden and 
Medical Psychology and Medical Sociology, Gutenberg-University, Mainz, Germany}\\

\noindent {\bf Abstract} {\small Graphical Markov models combine conditional independence constraints with 
graphical representations of stepwise data generating processes.
The models started to be formulated about 40 years ago and vigorous development is ongoing. Longitudinal  observational studies as well as intervention studies are best  modelled via a subclass called regression graph models and, especially  traceable regressions. Regression graphs include two types of  undirected graph and directed acyclic graphs in ordered  sequences of joint responses. Response components may correspond to discrete or continuous random variables or to both types and may depend exclusively on variables which have been generated earlier. These aspects are essential when causal hypothesis are the motivation for the planning of empirical studies.

To turn  the graphs into useful tools for tracing pathways of dependence, for understanding development over time  and for 
predicting structure in alternative models, the generated distributions have to 
mimic some properties of joint Gaussian distributions. Here, relevant results concerning these aspects are spelled out and illustrated by examples. With regression graph  models, it becomes feasible, for the first time,  to derive structural  effects  of (1) ignoring some of the variables,  of (2) selecting subpopulations via fixed levels of some other variables or of (3) changing the order in which the variables might get generated. Thus, the most important future  applications of these models  will aim at the best possible integration of knowledge from related studies.}\\
\noindent {\small {\bf Keywords} {\em Composition property, Conditional dependence, Conditional Independence, Connector transitivity, Directed acyclic graphs,
 Intersection property, Partial Closure, Partial Inversion, Regression graphs, Singleton transitivity, Traceable regressions, Undirected graphs.}}

\subsubsection*{Some historical remarks  and overview}
Graphical Markov models provide the most flexible tool for formulating, analyzing, and interpreting relations among many variables. The models combine and generalise  three different concepts developed about a century ago: (1) directed  graphs, in which variables are represented by nodes, used  to study linear processes by which joint distributions may have been generated (Sewell Wright, \cite{bibWright23, bibWright34}; \cite{bibTukey54}), (2) simplification of a joint distribution with the help of conditional independences (Andrei A. Markov, \cite{bibMarkov12}), and (3) specification of  associations only for variable pairs which are in some sense strongly related and are turned into nearest neighbors in an undirected  graph (Willard Gibbs, \cite{bibWGibbs02}; \cite{bibSpeed79}). 

First formulations of graphical Markov models started about 40 years ago, \cite{bibWer76a,bibWer76b,bibWer80}, \cite{bibDarLauSp80}, \cite{bibWerLau83}, several books with differing emphases 
have appeared since then, for instance, \cite{bibWhitt90}, \cite{bibOlivSmith90}, \cite{bibLau96}, \cite{bibCoxWer96}, \cite{bibEdwards00}, \cite{bibGreenHjortRich02}, \cite{bibStuden05}, \cite{bibWainJor08}, \cite{bibHosEdwLau12}.
Vigorous development is ongoing.  These  multivariate statistical 
models combine the above  simple but most powerful notions: data generating processes 
in sequences of single or of joint responses and conditional 
independences and dependences captured by graphs. Arguably, the most outstanding feature of
these types of models is that many of their implications can be derived using the graphs. Some of this 
will be outlined and illustrated here. 

The generating processes  concern no longer only linear relations, as a century ago, but they
include,  among others, linear regressions, \cite{bibWeisb14}, generalized linear models, \cite{bibMcCNel89}, \cite{bibAndSkov10}, exponential response models, \cite{bibHaberm77}, \cite{bibBarnd78},
subclasses of structural equations for longitudinal 
studies, \cite{bibjoresk81},  \cite{bibBollen89},  models for planned interventions such as  controlled clinical trials with randomized 
 allocation of individuals to treatments,
 and models for only virtual interventions, \cite{bibSpirGlySch93}, \cite{bibPearl09}, \cite{bibUhleretal}.
In particular, response variables may in general be vector variables that contain 
discrete or continuous variables or both types as  components. 

We concentrate here  on ordered  series 
of regressions for which the  responses have as  regressors exclusively  variables,
which have been generated earlier, so that they are  in the past of the response.  Throughout,  we use the terms  regression and  conditional distribution  interchangeably.
The generated distributions are called traceable regressions, \cite{bibWer12}, when different
pathways of development can be traced in a corresponding graph, called their 
regression graph,  \cite{bibWerSadeg12}.  Regression graphs  extend  graphs for multivariate regression,
\cite{bibCoxWer93}, which are one of four different types of the so-called
chain graphs introduced  in the literature, \cite{bibDrton09}, \cite{bibLauWer89}, \cite{bibFryd90},
\cite{bibAndMadPer01}.  

Each such graph may  represent a research hypothesis
on how data could have  been generated, \cite{bibWerLau90} so that we speak of the
starting or the `{\bf generating graph}'. When one starts with such a general type of graph, one ordering of the joint responses is 
taken as fixed and the properties of regression graphs, stated here in Propositions \ref{prop:gregp1} and \ref{prop:gregp2},  assure that  their graphical 
structures have an interpretation in terms of probability distributions.

Often the objective is to uncover graphical representations that 
lead to an understanding of the generating process for appropriately
collected data. 
Then for each such  study, the starting point is the available 
substantive knowledge. It
is used to decide on variables that are relevant in a given context and on their ordering into responses, intermediate and explanatory variables. Explanatory variables or regressors may for instance be treatments, intermediate outcomes, risks or variables available at   baseline, that is at the start of  the study. The last are  named context variables since they
capture features that are taken as given, of the study or of the study individuals.  

Well-fitting graphs are derived  by using a combination of information from the study design, from
statistical analyses that are used to decide on  conditional dependences and  independences,  from past empirical evidence and from theoretically postulated relations.  For detailed analyses in some studies, see \cite{bibWerSadeg12}, \cite{bibWerCox13}; links to further sizeable empirical studies are in an overview,  \cite{bibWerCox15}.

In the following, we do not discuss  fitting- or model search-procedures in detail.  Instead,
we describe first  models and graphs for few variables; especially  graphs
that are fully directed or that are undirected, becuase they had been developed  first and are now still  intensively studied, 
mainly  in the context  of Bayesian inference or in computer science.
We then proceed to regression graphs and models, to special binary distributions, to a summary and some open problems.
The main purpose here is to introduce concepts, especially the interplay between generating processes, graphs, factorizations of densities, edge matrices and matrix operators to partially modify graphs or matrices.   Simple examples illustrate some of the now available, unifying results.

\subsubsection*{Directed acyclic graphs and three  \sVs}

We start by introducing  some terms commonly used for graphs  in order to  discuss  the three key situations for directed graphs.
A `{\bf graph}' consists of a {\bf node set}, $N=\{1, \dots, d\}$,
and one or more edge sets. Nodes are also called vertices. Two distinct nodes are said 
to be `{\bf coupled}', or to be adjacent, if they are directly linked in the graph. Such a
link is named an `{\bf edge}'.  A `{\bf simple graph}' has  at most one edge for each node pair and has no node linked to itself.  A graph is `{\bf complete}'  if all its node pairs are coupled.

 A sequence of edges connecting distinct
nodes is  a `{\bf path}'.  By convention, the shortest type of path is an edge. A `{\bf directed graph}' has exclusively  arrows as edges; it  is `{\bf acyclic}' if it is impossible to return to any starting node by following a `{\bf direction-preserving path}' that is
a sequence of arrows pointing in the same direction.  Directed acyclic graphs are simple graphs
and  each $ij$-arrow, $i  \fla j$, points from  a regressor node  $j$ to its response node $i$; or are said to point from a parent $j$ to its child $i$.  We shorten the name `subgraph induced by a set of nodes', to  `{\bf subgraph of nodes}', which  just keeps those nodes and the edges present among them in a given graph.

\begin{figure}[h]
\centering
\includegraphics[scale=.54]{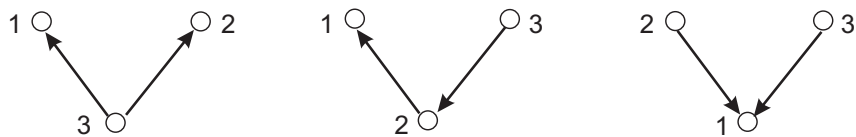} 
\caption{\textit{
The three types of \sV in  directed acyclic graphs; left: source \sV\!, middle:  transition \sV\!, right:  sink \sV\!, called in the literature also a collision \sV\!,  an unshielded collider or unmarried parents having a common child.} }
\label{fig:Vdag}
\end{figure}

Fig.\ref{fig:Vdag} shows the possible three types of \sV in directed acyclic graphs.
A subgraph of three nodes   is called `{\bf a \sV}\!' if it has two
edges. In each \sV\!, there are two `{\bf outer nodes}' that are both coupled
to one common neighbor, the `{\bf inner node}' of the \sV\!. The name of a \sV stems from  its type of inner node.  
 In  a self-explanatory way,  the \sVs in Fig.\ref{fig:Vdag} are  called,  a `{\bf source \sV}\!' on the left, a `{\bf transition \sV}\!' in the middle and a
`{\bf sink \sV}\!' on the right.The notion of inner nodes extends to $ij$-paths i more than three nodes.

For just three variables  and in a condensed notation for the
generated probability density functions, the  factorizations corresponding to Fig.\ref{fig:Vdag} are
$$ f_{123}= f_{1 \mid 3}f_{ 2 \mid 3}f_{3}, \nn \n   f_{123}= f_{1 \mid 2}f_{ 2 \mid 3}f_{3}, \nn \n
f_{123}=f_{1\mid 23} f_{2}f_{3}.$$
The implied  constraints are conditional independence of the
outer node pair given the inner node, both on the left and in the middle, and
marginal independence of the outer node pair, on the right. In the notation
introduced by Dawid, \cite{bibDawid79}, one writes these constraints equivalently as
$$ (f_{1|23}= f_{1|3})\Leftrightarrow 1\ci 2|3, \nn (f_{1|23}= f_{1|2})\Leftrightarrow 1\ci 3|2, \nn (f_{23}=f_{2} f_{3})\Leftrightarrow 2 \ci 3, $$
again in a condensed notation in which each node denotes also a variable.

Only the generating process in the middle of Fig.\ref{fig:Vdag} specifies a full
ordering of all three variables as $(1,2, 3)$, while  
one cannot distinguish with
the graph alone between $(1,2, 3)$ and $(2, 1, 3)$ for the source \sV and between
$(1, 2,  3)$ and  $(1, 3, 2)$ for the sink \sV\!\!. More generally, a directed acyclic graph
may  be `{\bf compatible with several orderings}'  of the variables such that the set
of all independences, that is the `{\bf independences structure}' of a graph, remains
unchanged. This poses problems for some machine-learning strategies.
In many applications however, one compatible ordering can  be taken as fixed;
 substantive knowledge may  even give a full ordering
of all  variables.

\subsubsection*{Parent graphs and three  \sVs}
A graph  is said
to form a `{\bf dependence base}'  if  a full ordering of the nodes is fixed and each edge present in the graph means the lack of  a conditional  independence, typically a dependence that is considered to be strong in a given context.   General properties of the graphs are also used.  For regression graphs, these are stated here in Propositions  \ref{prop:gregp1} and \ref{prop:gregp2}. Directed acyclic graphs that form a dependence base have been named  `{\bf parent
graphs}', \cite{bibMarWer09}, denoted by \Gpar\!.  Their defining pairwise relations are 
in equation \eqref{eq:defed}. 

For each node $i$ in the ordered node set, $N = (1, \ldots, d)$, of a parent graph,
one knows which nodes are in `{\bf the  past of node $\bm i$}', that is in set $\{>i\} = (i + 1, \ldots, d)$. The
subset of nodes in $\{>i\}$ from which arrows start and point to node $i$ is the set
of `{\bf parents of node $\bm i$}', denoted by $\pa_i$. In \Gpar\!, we have a dependence of
each node $i$ on all nodes in $\pa_i$ and independence of $i$ on all other
nodes in the past of $i$. Expressed by using the $\pitchfork$-notation introduced
for non-vanishing dependences by Wermuth and Sadeghi, \cite{bibWerSadeg12}, we have for  $j>i$ in \Gpar:
\begin{equation}i \pitchfork j |\pa_i\setminus \{j\}  \text{ for } j \in \pa_i \text{ and } i\ci j |\pa_i \text{ for } 
 j  \in \{>i\} \setminus \pa_i. \label{eq:defed} \end{equation}
 
As mentioned before, one outstanding feature of a graphical Markov
model is that its consequences can be derived, for instance for marginal or
for conditional distributions. To illustrate this  first for the graphs in
Fig.\ref{fig:Vdag}, we use a special notation.   A `{\bf boxed-in
node}',\,$\condnc$, indicates conditioning on the levels of the variable at this
node, and a `{\bf crossed-out}' node, $\margn$, means marginalizing over the variable, \cite{bibWerCox08}.

As justified later, we take  sink \sVs  in \Gpar to be {\bf edge-inducing} by conditioning and the source and transition \sVs\!, to be edge-inducing by marginalizing; each of the \sVs of Fig. \ref{fig:Vdag} introduces a different type of edge. The \sVs with the edge-inducing operation on the inner node  is shown in the following  line and the induced edges in the line thereafter.

\begin{equation}i \fla \margn \fra j , \nn \nn i \fla \margn \fla j, \nn \nn   i \fra \condnc \fla j \label{eq:Vpar} \end{equation}
$$   i\dal j,  \fourl  \fourl   \fourl i\fla j, \fourl  \fourl  \nn \nn i \ful j\,.$$
The induced edges `remember  at first'  the type of path ends at $i,j$ of the generating \sV, but then each  $\longleftrightarrow$  is replaced by $\dal$,  because no direction is implied after  ignoring a common source and, as explained below, the two types of undirected dependence can be readily distinguished. 

The following example is derived from information on a social survey, \cite{bibWainer83}. It
shows how conditioning  on the inner node of a  sink \sV\! induces a conditional dependence.  To distinguish underlying continuous variables from  discrete ones. The former are  drawn with a circle, the latter with a dot.
\begin{center}
 \includegraphics[scale=.56]{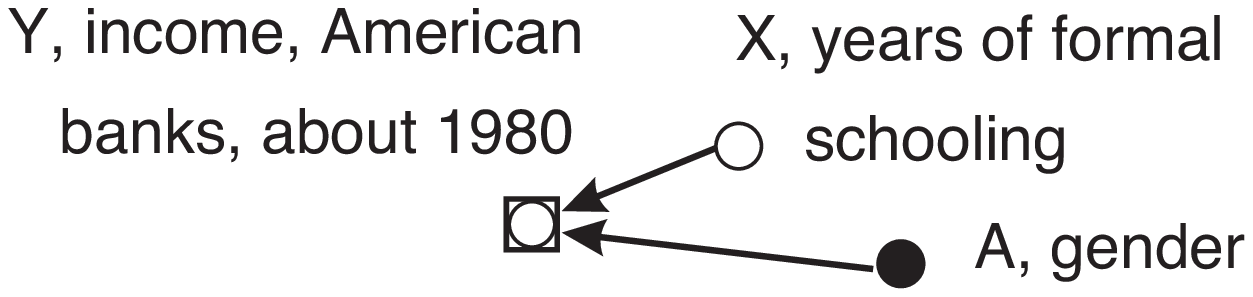}
\end{center}

In  American banks in the 1980s, salaries, $Y$, increased with  higher levels  of formal education, $X$, for both women and men,  that is $Y\pitchfork  X |A$, with $A$ denoting gender. Men received a clearly higher salary than women at given  levels of $X$, so that $Y\pitchfork A|X$. Furthermore, men and women had equal chances to obtain higher levels of formal education,  $X\ci A$.
This implies for $X \pitchfork A|Y$: for any  given  level of the salaries,
women had a higher level of formal education than men. 

 We show in the next section how  the above edge-inducing rules mimic the effects of marginalizing and conditioning in  non-degenerate Gaussian distributions,  those  that have invertible covariance matrices.

\subsubsection*{Gaussian distributions generated over parent graphs}

For linear relations in $d$ mean-centered variables $X_i$, a non-degenerate
Gaussian distribution is generated with
\begin{equation} {\bm{AX}} = {\bm \varepsilon}, \nn \E(\bm \varepsilon) ={\bm  0}, \nn \cov({\bm \varepsilon})={\bm \Delta} \text{ diagonal}, \label{eq:triang}\end{equation}
where zero-mean, uncorrelated Gaussian residuals, $\varepsilon_i$, have positive variances
$\sigma_{ii \mid >i }$ and are in the $d \times 1$ vector $\bm \varepsilon$. Vector $\bm X$ contains the variables  $X_i,$ and matrix
$\bm A$ is `{\bf unit upper-triangular}, that is it has ones along the diagonal and zeros below the diagonal. In row $i$, it has minus the values of linear regression coefficients resulting with
response $X_i$ regressed on $X_{>i}$, \cite{bibWeisb14}, \cite{bibWer80}.

In the early literature of econometrics, such linear relations have been discussed as recursive equations, \cite{bibStrotzWold60} and were written  in triangular form; for  three variables as:
\begin{eqnarray*} X_1 +a_{12} X_2 +a_{13}X_3&=&\varepsilon_1\,,\\
X_2 +a_{23}X_3&=& \varepsilon_2\,,\\
X_3&=& \varepsilon_3\,.
\end{eqnarray*}
 Note that in a Gaussian distribution generated over a complete parent graph, none of 
the regression coefficients vanishes when each response $X_i$ is regressed on all variables
in its past, that is on $X_{>i}$.

By equation (1),  missing
edges in the starting graph define the `{\bf independence constraints}'.
 For Gaussian distributions generated over parent graphs, these are  reflected  in vanishing regression coefficients and   
 as zeros in `{\bf matrices of equation  parameters}'.
For example, in the first and third generated distribution of Fig.\ref{fig:Vdag}, we can write:
$$ \begin{pmatrix} 1& 0& a_{13}\\ 0& 1& a_{23}\\0&0&1 \end{pmatrix}\begin{pmatrix} X_1\\X_2\\X_3\end{pmatrix}=
\begin{pmatrix}\varepsilon_1\\ \varepsilon_2\\ \varepsilon_3\end{pmatrix},\fourl 
 \begin{pmatrix} 1& a_{12}& a_{13}\\ 0& 1& 0\\0&0&1 \end{pmatrix}\begin{pmatrix} X_1\\X_2\\X_3\end{pmatrix}=
\begin{pmatrix}\varepsilon_1\\ \varepsilon_2\\ \varepsilon_3\end{pmatrix}\fourl $$
while for the second case in Fig.\ref{fig:Vdag}, $a_{12}$ and $a_{23}$  are nonzero but $a_{13} = 0$.

For an explicit distinction between conditional and marginal dependences,
we switch to a more detailed notation for trivariate Gaussian distributions.
For instance, $\beta_{1\mid 3.2}=-a_{13}$ is the coefficient of $X_3$ in the 
linear regression of $X_1 $ on $X_2$ and $X_3$, while $\beta_{2\mid 3} =- a_{23}$ 
is the coefficient of $X_3$  in the linear
regression of $X_2$ on $X_3$ alone. 

The  following relation between marginal and
conditional linear least-squares regression coefficients, due to William Cochran, \cite{bibCochran38}, is called the
recursion relation of these regression coefficients:
\begin{equation} \beta_{1\mid 3} = \beta_{1\mid 3.2} + \beta_{1\mid 2.3}\beta_{2\mid 3} \label{eq:rreg}\, . \end{equation}
Thus, for a Gaussian distribution generated over the parent graph in the middle of Fig.\ref{fig:Vdag},  which is a
transition \sV\!, 
 the conditional independence $1\ci 3|2$, $(\beta_{1|3.2}=0)$, implies the marginal
dependence $1\pitchfork 3$, $(\beta_{1|3}\neq 0)$,  because the edges present in the transition \sV mean $\beta_{1 \mid 2.3}\neq 0$ and $\beta_{2 \mid 3} \neq 0$.  This  property is shared by trivariate binary distributions, \cite{bibSimp51}. Joint distributions with this property in its generalized form, given here in equation \eqref{eq:singleton}, are said to be dependence inducing, \cite{bibWerCox98}, or to satisfy singleton transitivity,
\cite{bibWer12}.

For a Gaussian distribution generated over a \Gpar of Fig.\ref{fig:Vdag} on the right, which  is
a sink \sV\!,  the marginal independence $2\ci 3$ implies the conditional dependence $2\pitchfork  3|1$.
These features may best be recognized with equation \eqref{eq:indr} below, after  introducing correlations and their relations
to other types of parameter. 

With the covariance matrix denoted by $  \bm \Sigma$ and
its inverse, the concentration matrix, by $\bm \Sigma^{-1}$, we write explicitly
$$ {\bm \Sigma}= \begin{pmatrix} \sigma_{11}& \sigma_{12}&  \sigma_{13}\\
. &   \sigma_{22}& \sigma_{23}\\.&.&  \sigma_{33}\end{pmatrix}, \fourl
{\bm \Sigma}^{-1}= \begin{pmatrix} \sigma^{11}& \sigma^{12}&  \sigma^{13}\\
. &   \sigma^{22}& \sigma^{23}\\.&.&  \sigma^{33}\end{pmatrix} \,.$$
The $.$-notation indicates symmetric entries, the diagonal elements of $\bm \Sigma$ are
the `{\bf variances}', $\sigma_{ii}= \E(X^2_i)$, and the off-diagonal elements are the `{\bf covariances}',
$\sigma_{ij}= \E(X_i X_j)$,  of the mean-centered $X_i, X_j$. The diagonal elements of ${\bm \Sigma}^{-1}$ are
the  `{\bf precisions}', $\sigma^{ii}$,  the off-diagonal elements are the `{\bf concentrations}', $\sigma^{ij}$.

The `{\bf correlation coefficient}', $\rho_{23}$, and  the `{\bf partial correlation
coefficient}', $\rho_{23|1}$, relate to the other parameters 
and  to each other via
$$ \rho_{23}\!=\!\sigma_{23}/\sqrt{\sigma_{22}\sigma_{33}},\n
 \rho_{23\mid 1}\!=\!-\sigma^{23}/\sqrt{\sigma^{22}\sigma^{33}}\!=\! (\rho_{23} - \rho_{12} \rho_{13})/
 \sqrt{(1- \rho_{12}^2)(1- \rho_{13}^{2})},
 $$
$$\beta_{2\mid 3}=\!\sigma_{23}/\sigma_{33}=-\sigma^{23.1}/\sigma^{22.1}, \nn
\beta_{1\mid 3.2}=\!\sigma_{13|2}/\sigma_{33|2}=-\sigma^{13}/\sigma^{11}.$$
In this notation, $ \sigma^{23.1}$  is the concentration of $(2,3)$ after marginalizing
over $X_1$  and $\sigma_{13 \mid 2}$  is the covariance of $(1,3)$  conditionally given $X_2 = x_2$.

Correlations are best suited to reflect the strength of linear dependences, here those
induced by the independence constraints. With $2\ci 3$ and with $1\ci3 \mid 2$, the
 induced conditional and marginal dependences are, respectively,
\begin{equation} \rho_{23 \mid 1}*=- \rho_{12 \mid 3} \rho_{13 \mid 2}\,, \fourl
 \rho_{13}*= \rho_{12} \rho_{23}\, .\label{eq:indr}
\end{equation}  
Thus, the induced linear dependence can be considerably stronger for a marginal
than for a conditional independence. For instance with $2 \ci 3$, there is  $- \rho_{23 \mid 1} *>
0.96$ if $\rho_{12}= \rho_{13}=0.7$ and ${\bm \Sigma}^{-1}$ does not  exist if $\rho_{12}= \rho_{13}\geq \sqrt{0.5}.$
By contrast if $\rho_{12}= \rho_{23}=0.7$ and  $1\ci3|2$, the induced marginal correlation is only $\rho_{13}*=0.49.$

\subsubsection*{Some properties of Gaussian distributions}

There are recursions also for concentrations, \cite{bibDemp72}, and for covariances, \cite{bibAnd58}:
\begin{equation}  \sigma^{23.1}= \sigma^{23}-  \sigma^{12} \sigma^{13}/ \sigma^{11}, \fourl
\sigma_{13 \mid 2}= \sigma_{13}-   \sigma_{12} \sigma_{23}/ \sigma_{22}\, . \label{eq:rconcov}
\end{equation}
The first recursion shows that $0=  \sigma^{23.1}= \sigma^{23}$, that is both of ($2\ci 3$ and $(1\ci 2|3$) hold, if 
($\sigma^{12}=0$ or $ \sigma^{13}=0$) in addition.  Similarly, the second recursion shows
that both of ($1\ci 3|2$ and $1\ci 3$) hold  if ($ \sigma_{12}=0$ or $ \sigma_{23}=0$) in addition.
Thus,  an independence statement involving the third variable is needed for a variable
pair to be both marginally and conditionally independent.   This is the simplest case of  inducing dependences, that is of `{\bf singleton transitivity}'; see  \cite{bibWer12} and here equation \eqref{eq:singleton}.

Recursion relations such as  in equations \eqref{eq:rreg} and  \eqref{eq:rconcov} and their connection to the
elements of the above matrices $\bm A$ show also that in trivariate Gaussian distributions `{\bf conditional independences combine downwards}' as:
 $$ (1\ci 2\mid 3 \text{\:and\:} 1\ci 3\mid 2)\!\! \implies \!\!\!\{1\ci (2,3) \Leftrightarrow f_{123}\!=\!f_1 f_{23}\} \!\!\implies \!\!\! (1\ci 2 \text{\;and\;} 1\ci 3),$$
that is they satisfy what is also called the `{\bf intersection property}'. Furthermore, in these distributions
`{\bf conditional independences combine   upwards}' as:
 $$ (2\ci 3 \text{\;and\;} 1\ci 3)\!\! \implies \!\! \!\{3\ci (1,2) \Leftrightarrow f_{123}=f_{12} f_{3}\}\!\! \implies \!\!\!(2 \ci 3\mid 1 \text{\:and\:}1\ci 3\mid 2),$$ 
 that is they satisfy what is also called the `{\bf composition property}'.
  
In the information theory literature, non-degenerate Gaussian distributions have
been characterized by the above properties in terms of graphoids; these  structures  satisfy the  properties
common to all probability distributions plus  intersection, \cite{bibPearlPaz87}, \cite{bibStuden05}:
\begin{prop}\label{prop:1}{\em Ln\v{e}ni\v{c}ka and Mat\'u\v{s}, \cite{bibLnenMat07}.\!}
Gaussian distributions are singleton-transitive, compositional graphoids.
\end{prop}
 To make graphs useful tools for empirical studies, the  distributions generated over dependence-base 
 graphs have to share the properties of Prop.1 and are then called `{\bf traceable regressions}',
 \cite{bibWer12}; their graphs can be used to trace developmental pathways; see Example 2, given later.

Families of discrete distributions which violate singleton transitivity, 
the intersection or the composition property require
very special types of paramet\-rizations, \cite{bibWer12}. For the combination of independence statements of regression graphs,  the
intersection and the composition property are always used,  \cite{bibSadegLau14}. These
two properties also hold in  distributions generated
over  parent graphs; see  \cite{bibMarWer09}, discussion of Lemma 1, provided
the ordering and the dependences are indeed  as given with equation \eqref{eq:defed}. 
  
The  relations between linear parameters, discussed above, generalize to more than three
variables, but switching to a matrix notation and to edge matrix representations of graphs
becomes useful for  discussing most independence properties in general; for joint Gaussian distributions,  see for instance 
 \cite{bibMarWer09}, Appendix 2. Here, we
start again  with the  simplest type of edge matrices, those to the graphs of Fig.\ref{fig:Vdag}.

\subsubsection*{Structural versus parametric implications} 

 An edge matrix $\bm \acal$ can  be viewed as the sum
of an identity matrix, $\bm I$, and what has been named the adjacency matrix in graph theory; a square binary matrix with an $ij$-one if there is
a directed edge in the graph and an additional $ji$--one for an undirected edge. 
The small change of adding $\bm I$ leads to well-defined matrix products which can be used to derive  structural 
consequences of a given generating graph. As we shall see,  such structural consequences may differ from those of a given generating set of  parameters.

The edge matrices $\bm \acal$  in Table 1 share the unit upper-triangular form with
the linear equation  parameter matrices  $\bm A$ given with the generating  equations \eqref{eq:triang}.   \begin{table}[H]
\begin{center}
	\caption{\textit{Edge matrices  for the three \sVs  of Fig.\ref{fig:Vdag}}}
\vspace{1mm}
 \begin{tabular}{r c c c}
\hline \\[-4mm]
Edge matrices $\bm \acal$ of&$1\fla 3\fra 2$\n &\n $ 1\fla 2\fla 3$\n &\n $2\fra 1\fla 3$\\  \\[-4mm]
\hline \\[-3mm]
$\bm \acal:$&$\left(\begin{array}{rrr} 
                    1& 0&1 \\
                    &1&1\\
                   \bf 0&&1\\
                  \end{array} \right)$&
$\left(\begin{array}{rrr} 
                    1& 1&0 \\
                    &1&1\\
                   \bf 0&&1\\
                  \end{array} \right)$& 
$\left(\begin{array}{rrr} 
                    1& 1&1 \\
                    &1&0\\
                   \bf 0&&1\\
                  \end{array} \right)$\\  \\[-4mm]
\hline
\end{tabular}
\end{center}	
\end{table}

For instance with ${\bm M}\T$ denoting the transpose of a matrix $\bm M$, Gaussian systems, ${\bm{AX = \varepsilon}}$  of equation \eqref{eq:triang}, imply as covariance
and concentration matrices, 
\begin{equation} \bm{\Sigma= A}^{-1}\bm{\Delta}\, ({\bm A}^{-1})\T, \fourl  {\bm\Sigma}^{-1}= {\bm A}\T\, {\bm \Delta}^{-1}{\bm A} \label{eq:icovcon}\, , \end{equation}
where the matrix pairs $({\bm A}, {\bm \Delta}^{-1})$ and $({\bm A}^{-1}, {\bm \Delta})$ are  ordered Cholesky decompo\-sitions or
`{\bf triangular decompositions}' of ${\bm\Sigma}^{-1}$ and ${\bm\Sigma}$, respectively, \cite{bibWer80}.

For Gaussian distributions, zero elements in $\bm \Sigma$ and ${\bm \Sigma}^{-1}$ coincide with those  independences that hold, more generally,  in {\bf covariance and  concentration graphs}, respectively,  of other types of distribution:
$$ (\sigma_{ij}=0)  \Leftrightarrow i\ci j\,, \fourl (\sigma^{ij}=0)  \Leftrightarrow (i \ci j \mid N\setminus\{i,j\} ) \,.$$

For the dependence base \sVs  of Fig.\ref{fig:Vdag}, it may be checked directly that with $\Delta_{ii}=
\sigma_{ii \mid \pa_i}\!\!> 0$, a
nonzero element is induced  in different positions  in row one of ${\bm \Sigma}$ for the source \sV and for the
transition \sV\!, while a
nonzero element is induced in position $(2,3)$ of ${\bm \Sigma}^{-1}$  for the sink \sV\!;
see also equation \eqref{eq:indr}.

In general, implications of a graph result  via transformations of edge matrices. The edge matrix $\bm \acal$ of \Gpar, for  node set $N$ of size $d$, is 
the $d \times d$ unit upper-triangular matrix $\bm{\acal }= (\acal_{ij})$ such that
\begin{equation}
\label{eq:acal}
\acal_{ij} =
\begin{cases}
 1 & \text{ if  and only if } i \fla  j \text{ in } G^{N}_{\rm \pa} \text{ or } i = j, \\
 0 & \text{ otherwise.} \\
 \end{cases} 
\end{equation}

For  path interpretations, a definition of  node $j$ being an `{\bf ancestor}' of `{\bf descendant}'  $i$ is needed: there  starts a direction-preserving path at $j$ leading  to  node $i$.  We will now derive the edge matrix transformation that turns every ancestor in \Gpar into a parent.

The $k$'th power of the adjacency matrix  $(\bm{\acal- I})$ is known to count for each
$i < j$ in \Gpar  the number of direction-preserving paths of length $k$ connecting nodes  $i$ and $j$.
Since the longest of such paths has  $d-1$ edges, zero matrices  $(\bm{\acal- I})^k$ result
for all $k>d-1$.  Thus, the edge matrix of the ancestor graph of \Gpar, denoted by ${\bm{\acal}}^{-}$, becomes 
$$
{\bm \acal}^{-}\!=\!\In[(\bm{2I-\acal})^{-1}], \n (\bm{2I-\acal})^{-1}=\bm{I} + (\bm{\acal- I}) +(\bm{\acal- I})^2+\ldots +(\bm{\acal- I})^{\{d-1\}},
$$
where `In' is the indicator function that replaces every positive entry of a
nonnegative matrix by a one. The above sum is the matrix analogue to the
sum of an infinite geometric series, where for $ \mid a \mid< 1$, one obtains $(1-a)^{-1} =
1 + a + a^2 +  \dots$, (\cite{bibNeum1884}, p. 29, \cite{bibMabry99}).  This is generalized here in equation \eqref{eq:indcovcon}. The edge matrix analogue to equation  \eqref{eq:icovcon} is introduced next. 

With the edge matrices $\bm{\acal}$ and $\bm{\acal}^{-}$, the consequences  of  the starting graph, \Gpar\!,  for pairwise
marginal 
 and for conditional independences given all remaining variables,
 can be directly given. An implied independence $i\ci j$ and $i \ci j \mid N\setminus\{i,j\}$, respectively, is indicated by a zero in positions $(i,j)$ of 
\begin{equation}
{\bm \ncal}_{NN}=
 \In[\bm{\acal}^{-} (\bm{\acal}^{-})\T], \fourl   
 {\bm \ncal}^{NN}= 
 \In[\bm{\acal}\T \bm{\acal}]   \label{eq:indund}, \end{equation} that is  in the edge matrices of the `{\bf overall covariance and concentration graph induced by} {\bm \Gpar\!}'; see also equation 
 \eqref{eq:icovcon} and the next section.

Such zeros are said to be `{\bf structurally induced}' because they result for
all distributions that factorize as prescribed by a given generating graph.
With  the examples in Fig. \ref{fig:sink} and Fig. \ref{fig:sourtrans} in the next section, the types of path are identified which  induce more edges than there are present in a starting parent  graph and therefore lead to more complex structures, captured in one or both of the two  induced undirected graphs.

These  edge-inducing paths    introduce an additional   dependence in  Gaussian distributions generated over parent graphs,  provided  no other constraints apply than the  pairwise independences and dependences  defining their generating graph; see  equation \eqref{eq:defed} so that contributions of several paths may get cancelled.

If an independence is not structurally  induced, then it may still get generated by particular
constellations of the parameters. Such cases have been called `{\bf parametric
cancellation}', \cite{bibWerCox98}  or `{\bf lack of faithfulness to the
graph}', \cite{bibSpirGlySch93}.  For instance, a parametric
cancellation occurs if in equation \eqref{eq:triang}, one has  $\beta_{1\mid 3.2}=
-\beta_{1\mid 2.3}\beta_{2|3}.$ This leads to a zero in position $(1,3)$ of $\bm \Sigma$ even when $1\pitchfork 2 |3$ and 
$2\pitchfork 3$ and  hence to a non-structural independence, $1\ci 3$,
for Gaussian distributions.

\subsubsection*{Some consequences of a  five-node parent graph}

For five ordered nodes, $N=(1,\ldots,5)$, Fig.\ref{fig:sink} shows a parent graph, \Gpar,  which contains the three types of \sV 
of Fig.\ref{fig:Vdag}. Edges present and edges missing are defined by equation \eqref{eq:defed}. 
The factorization of $f_N$ can be read directly off  the graph:
$$ f_N=f_{1\mid 23}f_{2}f_{3\mid 5}f_{4\mid 5}f_{5}. $$ Also, the graph can be drawn using  the given order of the nodes and this factorization.

\begin{figure}[h]
\centering
\fourl  \fourl   \nn   generating parent graph   \hspace{1cm}  induced concentration  graph\\ \n \\
 \includegraphics[scale=.56]{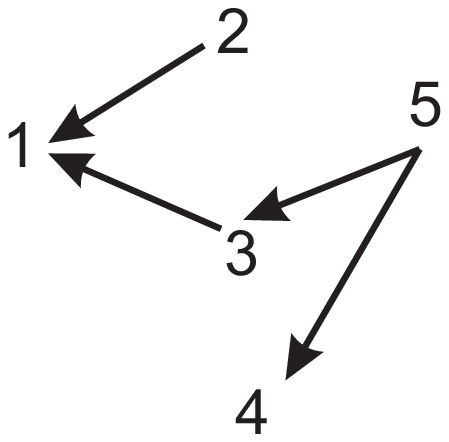}\nn \fourl   \fourl  \fourl  \fourl  \fourl \includegraphics[scale=.61]{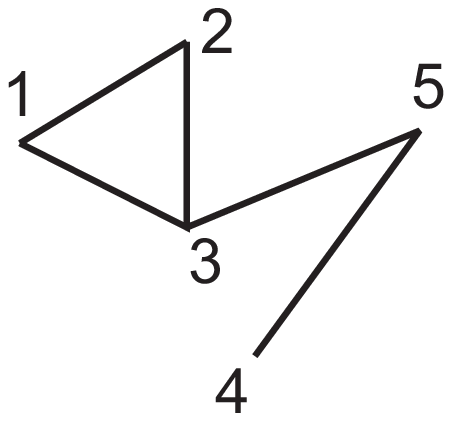}
\caption{\textit{left: a small  \Gpar  with the three  types of \sV\!,  with source node 5, transition node 3 and sink node 1; right: the induced concentration graph with an additional edge for $(2,3)$ due to conditioning on the common sink node 1 of (2,3).  
} }
\label{fig:sink}
\end{figure}

To generate the joint distribution over \Gpar,\! one starts with $f_5$, generates $f_{4\mid 5}$
next, then $f_{3\mid 5}$, then $f_2$ and finally $f_{1\mid 23}$.
The defining pairwise dependences in equation \eqref{eq:defed} give
$$ 1\pitchfork \{2,3\}\,,  \fourl 3 \pitchfork 5\, , \fourl 4 \pitchfork 5\,, $$
so that no simpler factorization holds in distributions generated over this parent graph.
From the  pairwise independences  in equation \eqref{eq:defed} or from  the factorization  of $f_N$, one obtains 
the defining  independence  structure of \Gpar as:
$$ 1\ci \{4,5\} \mid \{2,3\}\,, \fourl 2\ci \{3,4,5\} \, , \fourl  3\ci 4\mid 5.$$

All further implied independences may, in principle, be derived directly from such a  list
of  independences by using the properties of the starting  graph. We turn to these properties  in Prop. \ref{prop:gregp1},  \ref{prop:gregp2}. Similarly, further implied dependences
can be obtained by using the factorization of $f_N$ and the information that the factorization  cannot be further simplified. 
But, one may instead use the  edge-inducing properties, \cite{bibPearlWer94}, of  \sVs in  parent graphs, extending the discussion above for three-node graphs.
Proofs  may be  based on  Prop.  \ref{prop:4}, given later. We start with consequences of 
the sink \sV in Fig.\ref{fig:sink}. 

It can be derived  that for every sink {\sf V} 
with outer nodes $i,j$, all independence statements for $i$ and $j$ implied by \Gpar exclude
the inner sink node `o'. Here we have for instance, $2 \ci 3$, \, $2\ci 3|4$ and  $2\ci 3|\{4, 5\}$ so that there are several  subsets $c$ of $N\setminus \{i,o,j\}$ for which  $i \ci j|c$ is implied by the parent graph, here e.g. $c=\emptyset$, \, $c=\{4\}$, \, $c=\{4,5\}$. Thus, given a sink \sV\!, there are $c\subset N\setminus \{i,o,j\}$ such that $i\ci k|c$ is implied by the graph. For each such $c$,
\begin{equation} \text{ nodes } (i, {\rm o}, j) \text{ forming a sink \sV  in \Gpar}\Leftrightarrow (i\ci j|c \implies   i \pitchfork j|{\rm o} c)\,. \label{eq:sink}\end{equation}
In Fig. \ref{fig:sink}, for instance, $2\pitchfork 3|1$, $2\pitchfork 3|\{1,4\}$ and $2\pitchfork 3|\{1,4,5\}$
are induced. For Gaussian distributions, the size of such dependences can be expressed  in terms of induced partial correlations, in a similar way as in equation \eqref{eq:indr}.

The concentration graph induced by \Gpar\!,  involves conditioning on all nodes. The additional  edges  result by closing sink \sVs\!, as captured by  $\ncal^{NN}$ in equation \eqref{eq:indund}.
More edges represent  in general a more complex structure and in cases with complete,
undirected  subgraphs of three or more 
nodes, it cannot be recognized from a concentration graph alone which edges are due to conditioning on sink \sVs\! in \Gpar\!.  

We  turn next to consequences of
 transition and source \sVs by using Fig.\ref{fig:sourtrans}.  
It can be derived that for every  transition $\sf V$ 
with outer nodes $i,j$, all independence statements for $i$ and $j$ implied by \Gpar include
the inner node `o';. In Fig. \ref{fig:sourtrans}, we have for instance $1 \ci 5| 3$, $1\ci 5|\{2,3\}$,  so that there are several  subsets $c$ of $N\setminus \{i,o,j\}$ for which  $i \ci j|oc$ is implied by the parent graph, here such as  $c=\emptyset$, $c=\{2\}$.
\begin{figure}[h]
\centering
generating parent graph \n \fourl ancestor graph    \nn induced covariance graph\\ \n \\
  \includegraphics[scale=.56]{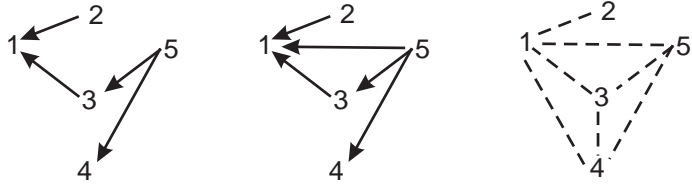}
\caption{\textit{left: the same generating  \Gpar as in Fig.\ref{fig:sink}; middle: the corresponding ancestor graph, also called the transitive closure of  \Gpar;
right: the induced covariance graph with new edges for $(1,4)$, $(1,5)$, $(3,4)$ compared to  \Gpar.}} 
\label{fig:sourtrans}
\end{figure}

 Thus, given a transition \sV\!, there are $c\subset N\setminus \{i,o,j\}$ such that $i\ci j|oc$ is implied by the graph. For each such $c$,
\begin{equation}\hspace{-3mm} \text{\;\,nodes } (i, {\rm o}, j) \text{ forming a transition \sV  in \Gpar}\! \Leftrightarrow \!(i\ci j|oc \implies   i \pitchfork j| c)\,. \label{eq:trans}\end{equation}
In Fig. \ref{fig:sourtrans}, for instance, $1\pitchfork 5$ and  $1\pitchfork 5|2$ are induced.
 A fully analogous statement results by replacing in the previous paragraphs each time  `transition node' by  `source node'. Just the examples relating to Fig. \ref{fig:sourtrans} change. 

The edge matrix $\ncal_{NN}$ in equation \eqref{eq:indund} shows that
by moving from the ancestor graph to the induced covariance graph, every source \sV in the former is closed by an edge.  Then, in the overall covariance graph induced  by \Gpar,  there is  an additional $ij$-edge if either $j$ is an ancestor of $i$ or $i$ and $j$ have a common ancestor.

Unless \Gpar  contains exclusively sink \sVs\!, there will be more edges
in the induced covariance graph. And again, whenever three or more nodes are contained in some of  its complete subgraphs,
it is impossible to see from the induced graph alone, whether additional dependences have been generated.
Therefore, the  same type of general warning as above applies to using the  class of covariance graphs for model  
selection. But furthermore,  when  a learning strategy is based on only the relations among variable pairs, no joint distribution may exist for such a given set of  two-way margins that results from joint distributions with higher-order interactions; for an example 
 see \cite{bibWerMar14}.  
 
To summarize, with information on the ordering of the variables, simpler structures will typically 
be uncovered, unless no additional edges are introduced, so that, say, a starting \Gpar and a concentration graph have  the same edge and node sets but different types of edge.
In such important  special  situations, there is   `{\bf Markov equivalence}', that is when the same independence structure is captured 
by two different graphs; see 
Prop. \ref{prop:5} below.

\subsubsection*{Undirected generating graphs}
Suppose now that   variables are unordered, that is arising at the same time, like several symptoms of a disease or local consequences of a global economic shock. 
Their joint distribution could then have a generating concentration graph,  \Gcon\!,  or a 
generating covariance graph, \Gcov\!. 
The defining pairwise independences for \Gcon are  $i\ci j|N\setminus\{i,j\}$ and those for \Gcov are
$i\ci j$. For  dependence base undirected graphs, each $ij$-edge present means:
 $$ i \ful j \Leftrightarrow i\pitchfork j|N\setminus\{i,j\} \text{ in \Gcon \nn and \nn } i\dal j  \Leftrightarrow i\pitchfork j \text{ in \Gcov} .$$

To read all implied independences off their graphs, a standard separation criterion from graph theory can be applied. For this, one says  `{\bf a   path intersects  a subset' } $g$} of node set $N$  if it has an inner node in $g$. We let  next $\{\alpha, \beta, c, m\}$ partition node set  $N$, where only $m$ or $c$ may be empty sets. This
notation is to remind one that with any independence statement $\alpha\ci \beta|c$, one implicitly has marginalised over the remaining nodes in $m=N\setminus\{\alpha \cup \beta \cup c\}$; one  considers the joint distribution of $Y_\alpha,Y_\beta$ given $Y_c$.
\begin{prop}\label{prop:2}{\em Darroch et al., \cite{bibDarLauSp80}}.
A  generating concentration graph, \Gcon\!, implies  $\alpha \ci \beta|c$ if every path  between $\!\alpha\!$ and $\!\beta\!$ intersects $c$.
\end{prop}

\begin{prop}\label{prop:3}{\em Kauermann, \cite{bibKau96}}.
A generating covariance graph, \Gcov\!,  implies  $\alpha \ci \beta|c$ if every path  between $\alpha$ and $\beta$ intersects $m$.
\end{prop}

Whenever these undirected generating graphs  also  form dependence bases,  the converse  holds as well: if a node in $\alpha$ is connected to one in $\beta$ by 
a path that does not intersect $c$ in \Gcon or that does not intersect $m$ in \Gcov\!, then  $\alpha \pitchfork \beta \mid c$ is implied.   Some additional effects  of Prop. \ref{prop:1} and  \ref{prop:2} result by considering  `$\bm a$-{\bf line $ij$-paths}': those which  connect node pair $i,j$  and have
all inner nodes  in   $a \subset N$.

\begin{coro} \label{coro:1} 
 By marginalizing over  any subset $a$ of $N$ in  \Gcon\!, all  $a$-line paths are closed 
while  by conditioning on $a$,  its subgraph  of $N\setminus a$ is  induced for $N\setminus a$.  By conditioning on subset $a$ in \Gcov\!,   all $a$-line paths 
are closed  while by marginalizing over $a$,  its subgraph of $N\setminus a$ is  induced  for $N\setminus a$. \end{coro}

Prop.  \ref{prop:1} and \ref{prop:2} imply more for `{\bf connected graphs}', that is when the nodes of every node pair can be reached via some path.

\begin{coro} \label{coro:2}\! A\! connected  \Gcon  induces for node set $N$ a complete covariance graph 
and a connected \Gcov induces for $N$ a complete concentration graph. 
\end{coro}

To summarize, in \Gcon\!, each full-line \sV is edge-inducing by marginalizing and in \Gcov\!, each dashed-line \sV is edge-inducing by  conditioning, where  we take again the induced edges to remember the  edge-ends of the  starting \sV\!. Note again that for Gaussian distributions generated  over a  dependence base \Gcon or \Gcov,
 an induced edge coincides always with an induced  dependence:
\begin{equation} \fourl i \ful \margn \ful j , \nn \nn i \dal \condnc \dal j, \fourl  \label{eq:Vund}\end{equation}
$$   i\ful j, \n \fourl  \fourl   \fourl  i\dal j\,.$$

The edge matrix of a complete generating graph of \Gcov  or \Gcon is a $d\times  d$ matrix of ones.
It has $d-1$ zero eigenvalues and one eigenvalue  equal to $d$. Hence, it is not invertible,
 but by subtracting it from a  $(d+1)$ multiple of an identity matrix,  one   obtains  a well-posed inversion task, \cite{bibTikhon63}.  In the statistical literature, this type of  {\bf \em Tikhonov regularization}  was introduced some fifteen years later in the form of  ridge regression;  a seemingly
 ill-posed problem is solved  by increasing  the diagonal elements of the matrix.
  
 If we denote by  ${\bm \wcal}$ any of  the symmetric edge matrices of a generating \Gcon or \Gcov\!,   then the corresponding edge matrices induced for  the covariance or the concentration graph are of the type:
  \begin{equation} {\bm \wcal}^{-}=  \In[\{(d+1){\bm \ical} -{\bm \wcal}\}^{-1}] . \label{eq:indcovcon} \end{equation} 
  By  definition, the  matrix $(d+1){\bm \ical} -{\bm \wcal}$ preserves the zero pattern of a  given edge matrix $\bm \wcal$  and it is a {\sf M}-matrix, so that its inverse is nonnegative.   The concept of a  {\sf M}(inkowski)-matrix was  introduced and studied by  Ostrowski,  \cite{bibOstro37} \cite{bibOstro56} without any applications concerning graphs or statistics;  it is an invertible matrix with exclusively nonpositive  off-diagonal elements.
For undirected generating graphs,  the {\sf M}-matrix in equation \eqref{eq:indcovcon} turns  each connected component  into a complete subgraph.

Figs. \ref{fig:sink} and \ref{fig:sourtrans} above illustrate, in particular, that a generating, undirected graph  is typically different from a corresponding induced graph. The latter summarizes all independences of a defined type implied by, say, a starting  \Gpar\!. It  can, in general, 
 not be used to derive  further implied independences of another  type; exceptions are discussed here later.  The concentration graph and the covariance graph induced by the \Gpar\!  in Figs. \ref{fig:sink} and \ref{fig:sourtrans} are  both incomplete, connected graphs. If they were also generating graphs, this would, by Corollary \ref{coro:2} or by equation \eqref{eq:indcovcon}, give a contradiction.
 
 In spite of the similarities of the two types of undirected graph, estimation of covariance graph structures, \cite{bibAnd73}, \cite{bibWerCoxMar06}, \cite{bibChaDrtRich07}, \cite{bibLupMarBer09}, \cite{bibKhareRaj11}, \cite{bibWonetal13}, is  typically much more complex than estimation of concentration
graph structures, \cite{bibDemp72},  \cite{bibWer80}, \cite{bibSpeedKiiv86}, \cite{bibCasRov06}, \cite{bibLnenMat07}). The latter but not the former have, for instance,  reduced sets of minimal sufficient statistics, \cite{bibBirch63},  \cite{bibCox06}, in exponential families with independences constraints, and for Gaussian distributions,  there is a unique maximum of the likelihood function whenever there are  less  variables
than observations (for `$p<n$'), \cite{bibDemp72}.

\subsubsection*{Regression graphs}
Regression graphs are simple graphs with response nodes in a set $u$ and 
context nodes in a set $v$ such that for an `{\bf ordered split}' of the node set as $N=(u,v)$, the density of the response vector, ${\bf X}_u$, is considered conditionally given the context  variables in vector ${\bf X}_v$
and
the joint density factorizes as
\begin{equation} f_N=f_{u\mid v} f_v\,. \label{eq:factcont}\end{equation}
Furthermore,  the response set $u$ has an ordered partition into connected components as  $u=(g_1, \ldots, g_k, \ldots, g_K)$ so that for all nodes in the subgraph of $g_k$, the nodes in their past 
are in $g_{>k}=\{g_{k+1},\dots, g_{K}, v\}$ and
\begin{equation} f_{u\mid v}=\txt \prod _{k=1}^{K} f_{g_{k}\mid g_{>k}}\label{eq:factresp}\,. \end{equation}
Simplifying  conditional independences are captured by  the  `{\bf 
regression  graph} {\bm \Greg\!}'. This simple graph uses Definition 1 and consists of  a concentration graph for the context nodes, a conditional 
covariance graph for  each of  the `{\bf concurrent responses}', that is for ${\bm X}_k$ within each connected component $g_{k}$, and a directed acyclic graph in the vector variables $({\bm X}_1, \ldots,  {\bm X}_K, {\bm X}_v)$.

 The  set-up for a regression graph model  starts with the response-vector variable ${\bm X}_1$ of primary interest,  possibly followed by one of secondary interest
and ends with a  context-vector variable ${\bm X}_v$;
 for an example see Fig. \ref{fig:setup}.
\begin{figure}[H] \vspace{-1mm}
\centering
 \includegraphics[scale=.55]{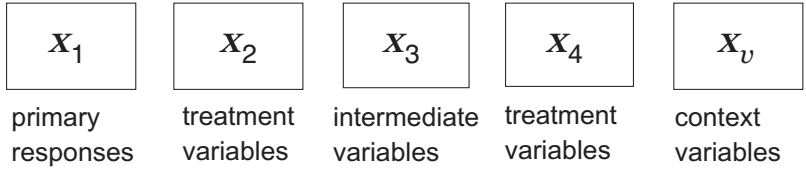}
\caption{\textit{A typical ordering of vector variables  for a regression graph model.}} 
\label{fig:setup}
\end{figure}
\noindent Intermediate variables
 form a sequence of  variables between ${\bm X}_1$ and  ${\bm X}_v$.

We let \Greg form a dependence base, so that edges present mean non-vanishing dependences; 
one ordering of the nodes is  fixed so that it is compatible with a known or hypothesized generating process.\\[3mm]
\noindent{{\bf Definition 1} Wermuth and Sadeghi,
\cite{bibWerSadeg12}.} \textit{An $ij$-edge present  in \Greg means} \\ 
$i \dal j$ with $i,j$ in $g_k$: \hspace{30mm} \fourl  \! ${i \pitchfork j| g_{>k}}$ ,\\
$i\fla j$ with $i$ in $g_k$ and  $j$ in  $ g_{>k}$: \nn  \hspace{13mm}\;${i\pitchfork j| g_{>k}\setminus \{j\}}$ ,\\
$i \ful j$ with $i,j$ in $v$: \hspace{29.5mm} \fourl \n $ { i \pitchfork j | v\setminus \{i,j\}},$\\[1mm]
\textit{while for uncoupled pairs $(i,j)$, the dependence sign $\pitchfork$ is replaced by the independence sign $ \ci$, but the conditioning sets remain unchanged}.\\[3mm]
There are equivalent pairwise properties of \Greg\!, \cite{bibSadWer15}, important
for interpretation, such as  $i \ci j |\pa_i$ for uncoupled $(i,j)$ with $i\in g_k$ and $j \in g_{>k}$.

A distribution is said to be  '{\bf generated over a regression graph}' when it satisfies
the factorizations of equations \eqref{eq:factcont},  \eqref{eq:factresp} while independences 
as well as dependences are specified by Definition 1 for a given node ordering $N=(1,\ldots, d)$.
Note that  with Definition 1,  \Greg is unchanged for a reordering of the nodes within any response set $g_k$.

In a regression graph,  three  additional \sVs may occur compared to those in a parent  graph, see equation \eqref{eq:Vpar}, and in the two types of undirected graph, see equation \eqref{eq:Vund}:
\begin{equation}i \fla \margn \dal j , \nn \nn i \fla \margn \ful j, \nn \nn   i \dal \condnc \fla j \label{eq:Vreg} \end{equation}
$$   i\dal j,  \fourl  \fourl   \fourl i\fla j, \fourl  \fourl  \nn \nn i \fla j\,.$$

With Definition 1  and a fixed compatible  ordering of the nodes,
three edge sets of different types are given for \Greg\!, in a self-explanatory notation, as
$  E_{\dals}\!,  E_{\flas},  E_{\fuls}$. Their union defines one edge set $E$. Three different  \sVs are edge-inducing by conditioning on the inner node, see the last \sV on the right-hand side of equations \eqref{eq:Vreg}, \eqref{eq:Vund}, \eqref{eq:Vpar}. These are the `{\bf collision \sVs}\!', the other five possible types  of a \Greg  are the `{\bf transmitting \sVs}\!'. Accordingly,  the inner nodes are  '{\bf collision nodes}'  or `{\bf  transmitting nodes}'.

One justification for the types of induced edge stems from the construction of summary 
graphs, one class of `{\bf independence-preserving}' graphs, those which preserve all independences implied by a generating
\Gpar or \Greg  in  a smaller graph obtained after marginalizing, conditioning and removing
nodes as well as  their edges,  \cite{bibWer11}, \cite{bibSadeg13}, \cite{bibSadegMar12}. In particular,  the above different 
types of  \sVs, plus two more that are used in constructing summary graphs,
can be  combined in any order in a consistent way, \cite{bibWer11} Appendix.

There are other classes of independence-preserving graphs,  not described here, by which additional implications of a generating graph may be derived from a smaller graph. These 
may have different types of edge,
\cite{bibKost02}, \cite{bibRichSpir02}, \cite{bibSadeg13}, serve different  purposes but define the same independence structures. 

To derive independences implied by \Greg\!, further concepts are useful. 
The notion of   anteriors of a response $i$, \cite{bibSadeg09}, extends the one of ancestors in \Gpar\! to \Greg\!.   For $N=(u,v)$
and  $i\fla j$,  node $i$ in $g_k$ is within $u$,  but  the parent node $j$ can be any node in $g_{>k}$,
the past of $i$. `{\bf Anterior paths}' join paths among context nodes in $v$ with an arrow to  descendant-ancestor 
paths in $u$:
 \begin{displaymath}
i \
\fla \underbrace{\overbrace{ \snode \fla \snode, \ldots,  \snode \fla  d_u}^{\text {\normalsize{ancestors of  \textit{i}}}}\ful  1\ful \snode, \ldots , \snode \ful d_v}_{\text{\normalsize{anteriors of \textit{i}}}}\,.
 \end{displaymath}
 Recall that an  $a$-line path connects a node pair  by a path with all inner nodes in  a subset
 $a$ of $N$. With this,   the notion of an ancestor graph of \Gpar can be extended.\\[3mm]
 \noindent {\bf Definition 2} \textit{An $\bm a$-{\bf line anterior graph} of \Greg  has edge\;$i\fla j$\;for every 
 $a$-line anterior $j$ of $i$ in \Greg and $a$-line paths for context nodes in $v$ are closed.}\\[3mm]
This graph permits to express  the effects of separation in \Greg\!,  \cite{bibSadeg09}, \cite{bibSadegLau14}, in a way comparable to those in undirected graphs, see Prop. \ref{prop:2} and
\ref{prop:3}. 
Again, let $N=(a,b)$, $a=\{\alpha, m\}$ and  $b=\{ \beta, c\}$,  where  only $m$ or $c$ may be empty.
\begin{prop} {\em Wermuth and Sadeghi, \cite{bibWerSadeg15}}. \label{prop:4} A regression graph implies $\alpha \ci \beta|c$ if  along every path 
between $\alpha$ and $\beta$, 
 in the $a$-line anterior graph of \Greg\!,  a collision node intersects $m$ or a transmitting node intersects $c$.
\end{prop}
 The converse of Prop. \ref{prop:4} holds with Definition 1, and a fixed   compatible ordering of the nodes. 
Prop. \ref{prop:4}  specializes to the effects of separation 
 in directed acyclic  graphs; see \cite{bibMarWer09}, Criterion 1, also for a   proof  of equivalence to other  path criteria for the independence implications of the \Gpar, \cite{bibPearl88}, \cite{bibGeiVerPea90}, \cite{bibLauetal90}. 
 
 \begin{coro} A path between $\alpha$ and $\beta$ in the $a$-line anterior  graph of \Greg  is  edge-inducing  if every collision node is in $c$ and every  other node is in $m$.\label{coro:3}
 \end{coro}
 \begin{coro} A path between $\alpha$ and $\beta$ in the $a$-line ancestor  graph of \Gpar is  edge-inducing  if every sink node is in $c$ and every  other node is in $m$. \label{coro:4}
 \end{coro}
 
 In particular, \Greg and \Gpar  induce a complete graph by marginalizing over $N$  if,  for node 1,  the last node $d$ is an anterior in \Greg or an ancestor in \Gpar.

  One  further important  question is whether two regression graphs with different types of edge can define the same independence structure
 if they have the same node set $N$ and an identical edge set $E= E_{\dals}\! \cup  E_{\flas}\cup  E_{\fuls}$. 
\begin{prop}{\em Wermuth and Sadeghi, \cite{bibWerSadeg12}}. \label{prop:5} Two regression graphs, with different types of edge but an identical node set  $N$ and an identical edge set $E$,   are Markov equivalent if and only if
their sets of collision \sVs coincide. 
\end{prop}
Thus for instance, a given regression graph is Markov equivalent to its induced concentration graph if and only if \Greg does not contain a collision \sV, and to its induced covariance graph if and only if \Greg does not 
contain any transmitting \sV\!. A covariance and a concentration graph are  Markov equivalent  if and only if they consist of identical sets of complete subgraphs.

Before we derive  graphs induced by \Greg\!,  we introduce  two basic types of Gaussian  distributions that may get generated over a regression graph.

\subsubsection*{Two types of Gaussian regression graph model}

  We note first that for  mean-centered variables and $N=(a,b)$,
a  `{\bf linear regression of a  joint response ${\bm X}_a$  on ${\bm X}_b$} gives, \cite{bibWeisb14},  
 \begin{equation}{\bm X}_a= {\bm \Pi}_{a|b}
 {\bm X}_b+{\bm \eta}_a, 
 \n \E({\bm \eta}_a)={\bm 0}, \n  \cov({\bm \eta}_a,  {\bm X}_b) ={\bm 0}, \n \cov({\bm \eta}_a) 
  \text{ invertible}.  \label{eq:mreg}\end{equation}
  The parameters are  a matrix of population least-squares
 regression coefficients ${\bm \Pi}_{a| b}$ and a residual covariance matrix $ {\bm \Sigma}_{aa|b} =\E({\bm \eta}_a {\bm \eta}_a\T)$. The interpretation  of ${\bm \Pi}
 _{a|b}$  results by post multiplication in the first equality of equation \eqref{eq:mreg} with ${\bm X}_{b}\T$ and taking expectations:
$ \E( {\bm X}_{a}{\bm X}_b\T)-{\bm \Pi}_{a|b}\E({\bm X}_{b}{\bm X}_b\T)\!\!=\!{\bm 0}.$ 

 Joint Gaussian distributions generated over a corresponding 
 regression graph are non-degenerate,  have a concentration matrix $\bm{\Sigma}^{bb.a}$ for  ${\bm X}_b$ and zeros 
 in  the defined parameter matrices are given by Definition 1 for $K=1$.
  
 The well-known relations of these parameter matrices, \cite{bibDemp69}, \cite{bibDemp72} Appendix B, \cite{bibMarWer09}\! Appendix \! 1, to   ${\bm \Sigma}=\cov({\bm X}_{N})$ and to ${\bm \Sigma}^{-1}$ are for  $N=(a,b)$:
 $$ {\bm \Sigma}=  \begin{pmatrix} {\bm \Sigma}_{aa} &{\bm \Sigma}_{ab} \\[1mm]
                           . & {\bm \Sigma}_{bb}\\
                         \end{pmatrix}, \nn \nn                  
                        {\bm  \Sigma}^{-1}=  \begin{pmatrix} {\bm \Sigma}^{aa} & {\bm \Sigma}^{ab} \\[1mm]
                           . & {\bm \Sigma}^{bb}
                         \end{pmatrix} ,$$\\[-12mm]
 \begin{eqnarray}
 {\bm \Sigma}_{aa|b}&= &{\bm \Sigma}_{aa}^{\n} -{\bm \Sigma}_{ab}^{\n} {\bm \Sigma}_{bb}^{-1}{\bm \Sigma}_{ba}^{\n}=({\bm \Sigma}^{aa})^{-1} \nonumber ,\\[1mm]
  \Pi_{a|b}&=&{\bm \Sigma}_{ab}^{\n}{\bm \Sigma}_{bb}^{-1}= -({\bm \Sigma}^{aa})^{-1}{\bm \Sigma}^{ab},  \label{eq:linp}\\[1mm]
  {\bm \Sigma}^{bb.a}&=&  {\bm \Sigma}^{bb}-{\bm \Sigma}^{ba}({\bm \Sigma}^{aa})^{-1}{\bm \Sigma}^{ab}={\bm \Sigma}_{bb}^{-1},\nonumber
 \end{eqnarray}
 where the expressions for   ${\bm \Sigma}^{bb.a}$ and ${\bm \Sigma}_{aa|b}$ are the matrix forms
of the recursion relations for concentrations and covariances in equation \eqref{eq:rconcov}. As is explained later, these
matrix results can all be obtained by applying the matrix operator named partial inversion.
The result analogous to Corollary \ref{coro:1} is the following  direct consequence of
equation \eqref{eq:linp}.

\begin{coro} \label{coro:3} 
 For any subset  $a$ of $N$,  marginalizing in ${\bm \Sigma}^{-1}$ over $a$ gives  ${\bm \Sigma}_{aa|b}$ and   ${\bm \Pi}_{a|b}$, while conditioning in ${\bm \Sigma}^{-1}$ on $a$ leaves the submatrix ${\bm \Sigma}^{bb}$
 unchanged.  For $b $ subset  of $N$, conditioning in ${\bm \Sigma}$ on $b$ gives $ {\bm \Sigma}^{bb.a}$ and $ \Pi_{a|b}$
 while marginalizing in ${\bm \Sigma}$ over $b$ leaves the submatrix ${\bm \Sigma}_{aa}$
 unchanged. 
 \end{coro}
This applies, in similar form also, for  $a=(\alpha, \gamma)$,   to  ${\bm \Sigma}_{\alpha\alpha|b}$ and with 
 $b=(\beta, \delta)$ to  ${\bm \Sigma}^{\beta\beta.a}$, that is marginalizing in any covariance matrix leads to a submatrix and conditioning in any concentration matrix leads to a submatrix, while more
non-vanishing parameters may get induced, otherwise.

For  $a=(\alpha, \gamma)$ and $b=(\beta, \delta)$, marginalizing over $\gamma$  and conditioning on  $\delta$  gives also a submatrix: ${\bm \Pi}_{\alpha| \beta. \delta}$,  where  $\alpha$ indicates the response, $\beta$ the  regressor and   $\delta$ the remaining regressors conditioned on:
\begin{equation} {\bm \Pi}_{a|b}= \begin{pmatrix} {\bm \Pi}_{\alpha|\beta.\delta} &{\bm \Pi}_{\alpha|\delta.\beta} \\[1mm]
                           {\bm \Pi}_{\gamma|\beta.\delta} & {\bm \Pi}_{\gamma|\delta.\beta}\\
                         \end{pmatrix} \label{eq:parmreg}.\end{equation}
Thus, by Corollary 5 and the same partition as in equation \eqref{eq:parmreg}, the parameters for  $f_{\alpha |\beta \delta}$ and $f_{\beta|\delta}$ are simply submatrices of those for $f_{a|b}$ and $f_{b}$.
 \begin{equation} {\bm \Sigma}_{\alpha\alpha |b}=[{\bm \Sigma}_{aa|b}]_{\alpha, \alpha}, \nn  {\bm  \Pi}_{\alpha|\beta.\delta}=[{\bm \Pi}_{a|b}]_{\alpha, \beta}, \nn \nn   {\bm \Sigma}^{\beta \beta.a}=[{\bm\Sigma}^{bb.a}]_{\beta, \beta}\,.\label{eq:submat}  \end{equation}

 The matrices  $ {\bm \Sigma}_{aa|b}$, ${\bm \Pi}_{a|b}$, $ {\bm \Sigma}^{bb.a}$ arise also, with   $\bm I$  denoting an identity matrix, in orthogonalized equations, \cite{bibWerCox04}, corresponding to equations \eqref{eq:mreg}:
\begin{equation} \begin{pmatrix} {\bm I}_{aa} & -{\bm \Pi}_{a|b}\\[1mm]
\bm 0&{\bm  \Sigma}_{bb}^{-1} \end{pmatrix} \begin{pmatrix}{\bm X}_a\\[1mm]{\bm X}_b\end{pmatrix}=
\begin{pmatrix} {\bm \eta}_a\\[1mm] {\bm \eta}_b\end{pmatrix}. \label{eq:orthjointr}\end{equation}
Note that $\cov ({\bm  \Sigma}_{bb}^{-1} {\bm X}_b)={\bm  \Sigma}_{bb}^{-1} $ so that this concentration matrix plays several roles.
 
For a {\bf Gaussian regression graph model}, recall that
 equations \eqref{eq:factcont}, \eqref{eq:factresp} and  Definition 1 apply. 
 Covariance matrices after regressing
${\bm X}_{k}$ on ${\bm X}_{>k}$ and ${\bm \Sigma}_{vv}^{-1}$ are in a block-diagonal matrix ${\bm W}_{NN}$. The matrix of equation parameters,   ${\bm H}_{NN}$, is  upper block-triangular  with identity matrices in the sizes of $g_k$ along the diagonal, ${\bm \Sigma}_{vv}^{-1}$ in the last block and off-diagonally $-{\bm \Pi}_{g_k\mid g_{>k}}$:
\begin{equation}{\bm H}_{NN}{\bm X}_N={\bm \eta}_N \text{ with }  {\bm W}_{NN}=\cov({\bm \eta}_N).\label{eq:regGpar} \end{equation}
As an example, we choose  $K=2$ and $u=(\alpha, \gamma)$:
$$ {\bm H}_{NN}=\begin{pmatrix} {\bm I}_{\alpha\alpha} & -{\bm \Pi}_{\alpha|\gamma.v} & -{\bm \Pi}_{\alpha | v.\gamma}\\[1mm]& {\bm I}_{\gamma\gamma} & -{\bm \Pi}_{\gamma| v}\\[2mm] {\bm 0}&  &  {\bm \Sigma}_{vv}^{-1} \end{pmatrix}\nn \nn  {\bm W}_{NN}=\begin{pmatrix} {\bm \Sigma}_{\alpha\alpha|\gamma v}& & {\bm 0}\\[1mm] & {\bm \Sigma}_{\gamma\gamma|v}& \\[2mm] {\bm 0}& & {\bm \Sigma}_{vv}^{-1} \end{pmatrix}.$$
Equation \eqref{eq:regGpar} implies for the single joint response regression of  ${\bm X}_{u}$  on  ${\bm X}_{v}$:
\begin{equation}{\bm P}_{u|v}=-{\bm H}_{uu}^{-1}{\bm H}_{uv},\nn  {\bm \Sigma}_{uu|v}= {\bm H}_{uu}^{-1} 
 {\bm W}_{uu}({\bm H}_{uu}^{-1})\T,\label{eq:uregv}
 \end{equation}
 hence with equations \eqref{eq:linp} also simple matrix expressions for $ {\bm \Sigma}^{uN}$ and  ${\bm \Sigma}_{Nv}$.\\[2mm]
The  edge sets of \Greg are captured by edge matrices ${\bm \hcal}_{NN}$ and ${\bm \wcal}_{NN}$.\\[3mm]
\noindent{\bf Definition 3}
We denote the 
dimension  of $g_k$   by  $d_k$, the one of $v$ by $d_v$, so that
$d=\txt{\sum}_{k=1}^{k=K} d_k  + d_v$ for the ordered node set $N=(1,\ldots, d)$.
The edge matrix,  $\bm \hcal=(\hcal_{ij})$,  is upper block-triangular, with  $K$ identity matrices of size $d_k \times d_k$
along the diagonal and a symmetric edge matrix for   the concentration graph of ${\bm X}_v$ alone
in the last block. In the upper, off-diagonal  parts  are ones for arrows pointing  in \Greg  from  $g_{>k}$ to $g_k$: 
\begin{equation}
\label{eq:hcal}
\hcal_{ij} =
\begin{cases}
 1 & \text{ if  and only if } i \fla  j \text{ or }   i \ful  j  \text{ in \Greg or } i = j, \\
 0 & \text{ otherwise.} \\
 \end{cases} 
\end{equation}

The edge matrix, $\bm \wcal_{uu}=(\wcal)_{ij}$, for dashed lines, is  block-diagonal with   $K$ 
symmetric $d_k \times d_k$ edge matrices for covariance graphs of ${\bm X}_k$ given ${\bm X}_{>k}$:
\begin{equation}
\label{eq:wcal}
\wcal_{ij} =
\begin{cases}
 1 & \text{ if  and only if } i \dal  j \text{ in \Greg or } i = j, \\
 0 & \text{ otherwise.} \\
 \end{cases} 
\end{equation}
For  ${\bm \wcal}_{NN}$, the last block is  taken to be
${\bm \wcal}_{vv}={\bm \ical}_{vv}$ because the full line edges of the concentration graph of ${\bm X}_u$ are already captured by ${\bm \hcal}_{vv}$.

For nodes $u$ as a single response node and $v$ its regressors, one gets e.g.
\begin{equation}{\bm \pcal}_{u|v}=\In[{\bm \hcal}_{uu}^{-}{\bm \hcal}_{uv}],\nn  {\bm \scal}_{uu|v}= \In[{\bm \hcal}_{uu}^{-} 
 {\bm \wcal}_{uu}({\bm \hcal}_{uu}^{-})\T],\label{eq:eduregv} \end{equation}
 as the  induced edge matrices for  equation \eqref{eq:uregv}. Note  that some induced edge matrices are denoted by using a close calligraphic equivalent
 to the parameters in the Gaussian case. These are  then either edge matrices of a starting graph, or their  submatrices, or they  can be derived directly
in terms of the matrix operator described next.  All others get  ${\bm \ncal}$ as notation. 

  \subsubsection*{Partial closure}
We now let the ordered node set $N=(1, \ldots, d)$ denote also the rows and columns of an edge matrix  $\bm \mcal$
containing \sVs of one type. Then, for an ordered partitioning as $N=(a,b)$,  the `{\bf partial closure}' operator, denoted by $\zer_a {\bm \mcal}$, closes $a$-line paths in a corresponding graph with edge matrix ${\bm{\mcal}}$
 and finds structural zeros induced in a  corresponding  parameter matrix $\bm M$  of a Gaussian distribution.\\[2mm]
\noindent{{\bf Definition 4}    The partial closure operator is, with $a=\{1\}$, 
$$ {\bm \mcal}= \begin{pmatrix}  1 & {\bf v}\T\\ {\bf w}& \bf {m}\end{pmatrix}: \nn  \n \zer_{\{1\}}\, {\bm \mcal}=\begin{pmatrix} 1  & \bf{v}\T\\ {\bf w}& \In[\bf{m}+\bf{wv}\T]\end{pmatrix}, \label{pinvd}$$ 
and   $\zer_a  {\bm \mcal}$, for $t>1$ elements in $a$,  may be thought of as applying the above  operation $t$ times,  using repeatedly appropriate permutations of $\bm \mcal$.\\[2mm]
An off-diagonal $i,j \neq k$  of  $\zer_{\{k\}} {\bm \mcal}$ contains an additional one compared to ${\bm \mcal}$  if and only if $\mcal_{ij}=0$ and $\mcal_{ik}\mcal_{kj}=1$, hence indicating the presence of a \sV in the graph. The operator preserves all ones of $\bm \mcal$ and it closes paths with \sVs which must be of the same type in the graph represented by $\bm \mcal $.

\begin{prop}\label{prop:6}{\em Wermuth, Wiedenbeck and Cox, \cite{bibWerWieCox06}}.
Partial closure is commutative, cannot be undone and is exchangeable with taking 
submatrices. \end{prop}
One may for instance get the  edge matrix ${\bm \acal}^{-}\!$, that is obtain the transitive closure of a directed acyclic graph,   and ${\bm \wcal}^{-}\!$ of equation \eqref{eq:indcovcon},  that is  complete all connected components  of an undirected graph, with      $N\!=\!\{a,b\}$ as
$$\zer_b \zer_a {\bm \acal}=\zer_N {\bm \acal} ={\bm \acal}^{-}, \fourl \zer_b \zer_a {\bm \wcal} =\zer_N {\bm \wcal} ={\bm \wcal}^{-}.$$
By Definition 1 and Prop. \ref{prop:6},
 $\zer_a \bm \acal$  for \Gpar remains   unit-upper triangular in the starting order, $\zer_a  \bm \wcal$ for \Gcon remains symmetric and disconnected 
 components, such as the graphs for conditional covariances of different joint responses in \Greg\!, remain disconnected. 

The edge matrices (${\bm \pcal}_{u|v}, {\bm \scal}_{uu|v}$)  in equation  \eqref{eq:eduregv},  induced by \Greg for $f_{u|v}$ with a
single joint response $u$,  may 
be obtained with $ [\zer_u {\bm \hcal}_{NN}]_{u,N}.$  For  $a=(\alpha, \gamma)$ and $b=(\beta, \delta)$, the edge matrix components for 
$ f_{\alpha |\beta \delta}$ and $f_{\beta|\delta}$ as induced by \Greg with $f_{a|b}$ and $f_b$ are given by the subgraph of $\alpha \cup \beta$,
just as the Gaussian parameters in equation \eqref{eq:submat} are given by submatrices. 
 
Algorithms  for finding  the  transitive closure in  directed graphs, possibly containing cycles, started to be developed independently in the Russian, French and American computer science literature; for a recent survey see \cite{bibVanSchaik10}. Algorithms for finding  connected components
for general graphs, \cite{bibTarjan72}, are  also still being developed, \cite{bibReing08}. 

One advantage of partial closure is
that its properties justify stepwise procedures using just the  \sVs in a \Greg\!. Another is that properties of this matrix operator prove some features of  the regression graph transformations.

  \subsubsection*{Edge matrices induced by {\bm \Greg}}
 
 The edge matrix of the $a$-line anterior graph of \Greg\!, see Definition 2, 
 arises for $a$ any subset of $N$, $b=N\setminus a$ and a reordering as $N=(a,b)$ with:
\begin{equation} {\bm \kcal}_{NN}=\zer_a {\bm \hcal}_{NN}. \label{eq:KNN}\end{equation}
This operation closes all full, $a$-line paths within $v$  and for each $i$ in 
$u$, it turns every $a$-line anterior $j$ into a parent of $i$. Similarly, 
\begin{equation}
{\bm \vcal}_{NN}=\zer_b{\bm \wcal}_{NN}, \label{eq:VNN} \end{equation}
closes all dashed, $b$-line paths in the conditional covariance graphs of the responses. 

 Thus, the two partial closure operations in equations \eqref{eq:KNN}, \eqref{eq:VNN} close all of the following four types of \sVs\!,  where ${\rm o}_g$ denote nodes  in a subset $g$ of $N$:
$$ i_u \fla {\rm o}_a  \fla  j_N, \n \nn i_u \fla {\rm o}_a  \ful  j_v, \nn \n i_v \ful {\rm o}_a  \ful  j_v, \nn \n i_u \dal {\rm o}_b  \dal  j_u\, ,$$
and the types of induced edge are as  specified in equations \eqref{eq:Vpar},  \eqref{eq:Vund}, \eqref{eq:Vreg} for ${\rm o}_a$, a node to be marginalized over, and ${\rm o}_b$, a node to be conditioned on.  These induced edges preserve the ordered split of the nodes, $N=(u,v)$. The corresponding model may be interpreted as a covering model, one with fewer constraints than 
the reduced model specified by the generating \Greg\!, \cite{bibCoxWer90}.

Four  types of \sV remain to be closed for  consequences of \Greg with $f_{u|v}f_v$ for $f_{a|b} f_b$:
$$ i_a \fla {\rm o}_a  \dal  j_a, \n \nn i_a \fla {\rm o}_a \fra j_b\, \nn \n    i_a \dal {\rm o}_b  \fla  j_b, \nn \n i_b \fra {\rm o}_b  \fla  j_b\, .$$
To achieve this,
 ${\bm \vcal}_{uu}$  is combined with  $ {\bm \kcal}_{vv}$ to give ${\bm \qcal}_{NN}$:
\begin{equation} {\bm \qcal}_{uu}= {\bm \vcal}_{uu},\nn {\bm \qcal}_{vv}={\bm \kcal}_{vv},\nn {\bm \qcal}_{uv}={\bm 0}\nn, {\bm \qcal}_{vu}={\bm 0}
 \label{eq:QNN}.\end{equation}
Then, these remaining  \sVs are closed  with the following  edge matrix products:
\begin{eqnarray*}
\In[{\bm \kcal}_{aa}{\bm \qcal}_{aa}{\bm \kcal}_{aa}\T]&\text{\!\!\!\!\! gives for}\!\!\!& i_a \fla k_a \dal  l_a  \fra  j_a \text{ and } i_a \fla k_a \ful  l_a  \fra  j_a\\
 & \fourl \nn \n& \text{a complete covariance graph} ,\\
 \In[{\bm \kcal}_{aa}{\bm {\bm \vcal}}_{ab}{\bm \kcal}_{bb}] &\text {\!\!\!\!\! gives for}\!\!\!& i_a \fla k_a \dal  l_b  \fla j_b
 \text{ a complete graph of response} \\
 & \fourl \nn \n & \text{nodes }  \{i_a, k_a\} \text{ and  regressor nodes } \{l_b, j_b\},\\
 \In[{\bm \hcal}_{bb}\T{\bm \vcal}_{bb}{\bm \hcal}_{bb}] & \text{\!\!\!\!\! \!gives for}\!\!\!&
 i_b\fra k_b  \dal  l_b  \fla j_b \text{ a complete concentration  graph.}
 \end{eqnarray*}

 This leads to  the edge matrix components,  ${\bm \ncal}_{a|b},\n {\bm \ncal}_{\,aa|b}$, of the graph  for regressing ${\bm X}_a$  on  ${\bm X}_b$ and,  ${\bm \ncal}^{\,bb.a}$, for the concentration graph of  ${\bm X}_b$, as induced by  \Greg\!;  induced arrows point from regressor  ${\bm X}_b$ to response ${\bm X}_a$.
 \begin{prop}\label{prop:7} {\em \! Wermuth, \cite{bibWer12}}\!. {Edge matrix components 
 induced by \Greg for $N=(a,b)$, by marginalizing over any $a \subset N$, conditioning on $b=N\setminus a$, are}
\begin{eqnarray*}{\bm \ncal}_{aa|b}&=&\In[{\bm \kcal}_{aa}{\bm \qcal}_{aa}{\bm \kcal}_{aa}\T], \\[1mm]
  {\bm \ncal}_{a|b}&=&\In[{\bm \kcal}_{ab}+{\bm \kcal}_{aa}{\bm \vcal}_{ab}{\bm \kcal}_{bb}],\\[1mm]
   {\bm \ncal}^{\,bb.a} &=&\In[{\bm {\bm \hcal}}_{bb}\T{\bm \vcal}_{bb}{\bm {\bm \hcal}}_{bb}].\end{eqnarray*}
    \end{prop}
   The zeros in these induced edge matrices represent  also  the structural zeros in $ {\bm \Sigma}_{aa|b}$, ${\bm \Pi}_{a|b}$, $ {\bm \Sigma}^{bb.a}$. 
As in Definition 1 for $K=1$, these $ij$-zeros mean:
 $$
  i \ci j| b   \text{ in } {\bm \ncal}_{aa|b}, \nn
  i \ci j|b\setminus j  \text{ in }  {\bm \ncal}_{a|b} , \nn
      i\ci j| b\setminus \{i,j\} \text{ in } {\bm \ncal}^{\,bb.a}.\\[3mm]$$
  \noindent {\bf Example 1} For \Gpar of Fig. \ref{fig:ex1} with edge matrix ${\bm \acal}$,  marginalizing with  an order-respecting split, $a=\{1,2,3\} $, and conditioning on $b=N\setminus a$  gives   
with ${\bm \acal}_{aa}^{-}\!=\!{\bm \kcal}_{aa}$ and ${\bm \kcal}_{ab}\!=\In[{\bm \acal}_{aa}^{-}{\bm \acal}_{ab}]$ a direct generalization of equation \eqref{eq:indund}:\\
$${\bm \ncal_{aa|b}}=\In[{\bm \acal}_{aa}^{-}({\bm \acal}_{aa}^{-})\T], \nn {\bm \ncal}_{a|b}={\bm \kcal}_{ab}, \nn {\bm \ncal}^{bb.a}=\In[{\bm \acal}_{bb}\T {\bm \acal}_{bb}]. $$  
     \begin{figure}[H]
\centering
\vspace{-2mm}
\includegraphics[scale=0.56]{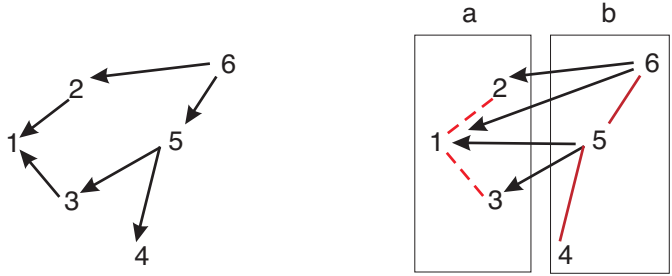}
\caption{\textit{Left: the generating parent graph, right:  induced graph for one set of response nodes,  $a=\{1,2,3\}$,  and one set of regressor nodes  $b=\{4,5,6\}$.}}
\label{fig:ex1}
\end{figure}
\noindent An alternative  to the edge matrix results is to use Corollary \ref{coro:4} to derive, separately for each missing edge in \Gpar,  the consequences  of the  new conditioning sets specified  by $N=(a,b)$ for  the regression graph of Fig. \ref{fig:ex1} on the right, which has only  one joint response, the one of nodes 1,2,3.  \\[3mm] 
 \noindent {\bf Example 2} A generating \Greg  for determinants of the well-being of diabetic patients having a lower level of formal schooling, $Y$,   is  given in Fig. \ref{fig:Ex2} left. The  following description of this graph attaches to it a  plausible, substantive  story. This uses  statistical results,  \cite{bibCoxWer96}, not given here. 
  \begin{figure}[H]
\centering   
\includegraphics[scale=0.56]{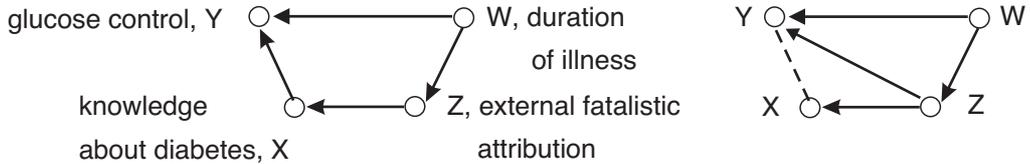}
\caption{\textit{Left: a generating \Gpar, right: the induced graph for regression of $(Y,X)$ on $(W,Z)$;  $Y\fla Z$  induced by implicitly marginalizing over X in  \Gpar to obtain the joint response $(Y,X)$.}}
\label{fig:Ex2}
\end{figure}

 Glucose control improves, the more  a patient knows about diabetes and the longer ago diabetes was diagnosed. Thus,
  glucose control depends
 directly on the knowledge about the illness, $X$, and on the time since the illness was diagnosed, $W$,
 hence $Y \pitchfork X| W$ and  $Y \pitchfork W| X$. Knowledge, $X$, is better,
 the lower the external fatalistic attribution, $Z$, that is the less  patients tend to think that their well-being
 depends mainly on their physicians, so that $X \pitchfork Z$. And, fatalistic attribution, $Z$, decreases with the time since diagnosis, $W$, so that $Z\pitchfork W$. This well-fitting graph contains the direction-preserving path $(Y,X,Z,W)$. This path, together with the  type of involved dependences,   suggests that intervening on the variables along it may improve the well-being of diabetic patients.\\[3mm]
 By constructing  induced graphs, one
 can answer queries like: which additional  dependences result 
 from a  given generating process by using another type of process for the same variables?
 This may  for instance arise in empirical studies when researchers disagree on the ordering of the variables. 
 In the example with $X$ as primary and $Y$ as secondary response,  $ X\fla W$ would result  due to conditioning on $Y$ in  the sink \sV, $(X,Y, W)$,  in the starting graph, while  for $Y,X$ as a joint response in Fig. 6 on the right, the   arrow $Y\fla Z$ is added due to marginalizing over the inner node $X$ of the 
 transition  \sV, $(Y,X, Z)$,  in the starting graph

 \subsubsection*{Edge criteria for effects of separation in  {\bm \Greg}} 

By Corollary 1 
and by ${\bm \ncal}_{a|b}$  representing the edge matrix induced by \Greg for the bipartite graph of arrows when ${\bm X}_a$ is regressed on ${\bm X}_b$,  submatrices of the edge matrices in Prop. \ref{prop:7} give also the structural zeros induced by \Greg for the joint conditional distribution 
with density of $f_{\alpha \beta |c}=f_{\alpha|\beta c} f_{\alpha|c}$: 
$$  {\bm \ncal}_{\alpha|\beta.c}=[{\bm \ncal}_{a|b}]_{\alpha, \beta}, \nn \nn   {\bm \ncal}_{\alpha\alpha |b}=[{\bm \ncal}_{aa|b}]_{\alpha, \alpha}, \nn \nn   {\bm \ncal}^{\beta \beta.a}=[{\bm \ncal}^{bb.a}]_{\beta, \beta}.$$

\begin{prop}\label{prop:10} {\em Wermuth, \cite{bibWer12}.}  A regression graph \Greg with edges given by Definition 1
  implies $\alpha\ci \beta|c$ if  ${\bm \ncal}_{\alpha|\beta.c}\!=\!0$ and it implies $\alpha\pitchfork \beta|c$ if  ${\bm \ncal}_{\alpha|\beta.c}\!\neq \!0. $
\end{prop} 

Thus, the absence of ones in a matrix indicates directly a queried independence and the presence of ones shows where dependencies occur.
Instead, with any path criterion, one has to study the properties of paths before a decision can be reached.  This may get  cumbersome
in large graphs when one has to check for each collision \sV whether its collision node is within the anterior set of $c$.

  \subsubsection*{Properties of regression graphs}
 
A regression graph, \Greg\!, shares the three properties in Prop. \ref{prop:1} of  a joint Gaussian distribution
generated over \Greg\!. It is dependence-inducing and independences combine downwards and upwards,
that is it  satisfies singleton transitivity, intersection and   composition, in addition to  the general   properties 
 of all probability distributions.
 
Its composition and  intersection property  have been proven in general, \cite{bibSadegLau14},
and were discussed above for just three variables. 
Singleton transitivity requires  an additional independence involving node $h$, say, if the conditioning 
 set for independence of  $i,j$  includes and excludes $h$.  For $i,j,h$ distinct nodes of $N$ and $c$ 
  a  subset of $N\setminus\{i,j,h\}$
\begin{equation} (i\ci j|c   \text{ and } i\ci j| hc) \implies (i \ci h| c \text{ or } j\ci h | c). \label{eq:singleton}\end{equation}
The  equivalent statement `for ($i \pitchfork h| c$  and $ j\pitchfork h | c$), either $i\ci j|c$ can  hold or  $i\ci j| hc$ but not both',  was proven with equations
\eqref{eq:sink} and \eqref{eq:trans} for two types of \sV in  \Gpar.\!  The same types of argument prove  singleton transitivity of \Greg\!. Recall that traceable regressions
 satisfy these same three properties, hence `{\bf mimic independence properties of a Gaussian distribution}' generated over \Greg.

For `{\bf set transitivity}' as defined in the literature, the single node $h$ in equation \eqref{eq:singleton} is replaced  by a  subset of $N$; disjoint of $\{i,j,c\}$.
Set transitivity is for instance violated when both of the  independence structures hold which are  defined  with the concentration and with the  covariance graph in Fig. \ref{fig:8}.

This may happen  in Gaussian distributions,   \cite{bibCoxWer93},  \cite{bibWer12}, but  not for undirected graphs; 
see Corollary 2.
More generally,  since graphs induced by \Greg may be  derived by partial closure and by adding products of binary matrices, 
contributions of several paths to a conditional dependence of any node pair $i,j$ can never cancel out. 
\begin{figure}[H]
\centering
\includegraphics[scale=.50]{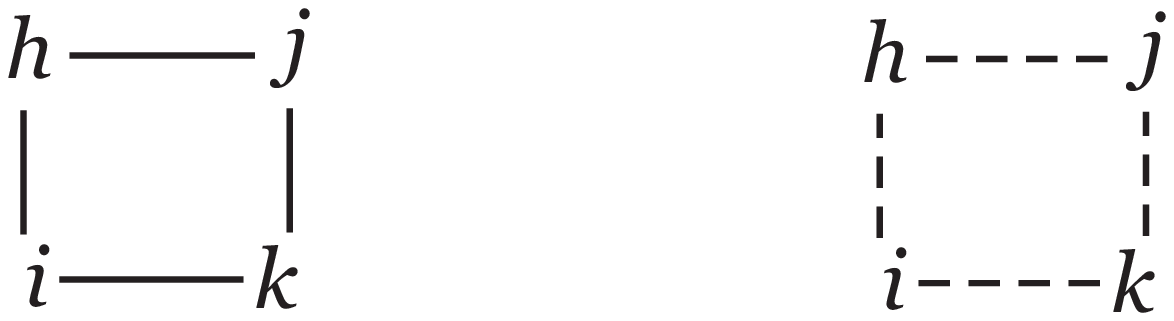}
\caption{\textit{Left: concentration graph with $i\ci j |\{ h,k\}$ and  $h\ci k |\{ i,j\}$; right: covariance graph with  $i\ci j$ and $h\ci k$;
connector for $i,j$ is $\{h,k\}$ in both.}} \label{fig:8}
\end{figure}
\noindent 
\begin{prop} {\em Wermuth and Sadeghi, \cite{bibWerSadeg15}}. The  structures captured and induced by \Greg are  like  traceable regressions but having and inducing exclusively positive dependences. \label{prop:gregp1}
 \end{prop}

 We show  next how
source, transition and sink \sVs of \Gpar  in Fig. \ref{fig:Vdag} and equation \eqref{eq:Vpar} generalize to source, transition and sink \sUs  in Fig. \ref{fig:Ufig}.
By remembering the path ends for the four $ij$-paths in Fig. \ref{fig:Ufig}, induced are  either  a dashed $ij$-line,  an $ij$-arrow or a full $ij$-line; see equations
 \eqref{eq:Vpar},  \eqref{eq:Vund}, \eqref{eq:Vreg} for the involved, repeated  closing of \sVs\!:
 \begin{figure}[H]
\centering
 \includegraphics[scale=.49]{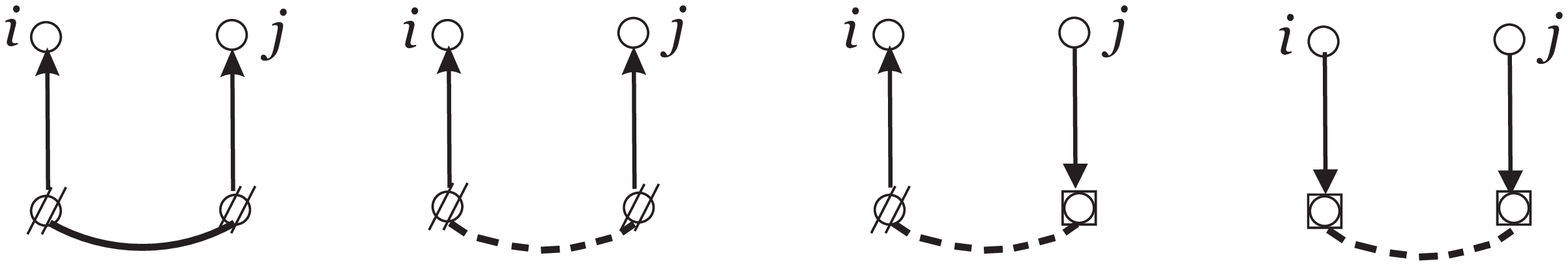}
\caption{\textit{Types of \sU with undirected edges and arrows to or from $i,j$. The first two on the left:  source \sU\!; the third: transition \sU\!;  the fourth on the right: sink \sU\!.}}  
\label{fig:Ufig}
\end{figure}

We now let $i$ and $j$ be again an uncoupled node pair  of $N$. Sets $\delta\neq \emptyset$ and $c$ be disjoint subsets of $N\setminus \{i,j\}$. Then, $\delta$ is  called a  `{\bf connector}' if  \Greg implies  ($i\ci j|\delta c $ and $i\pitchfork  j|c$)  or   ($i\ci j|c $ and $i\pitchfork  j|\delta c$) and the inner nodes of undirected $ij$-paths exhaust the nodes of $\delta$. With this definition, a previous claim of set transitivity of \Greg, \cite{bibWer12}, can be corrected as follows. 

\begin{prop} {\em Wermuth and Sadeghi, \cite{bibWerSadeg15}}. Regression graphs are connector-transitive, compositional graphoids.
\label{prop:gregp2}
 \end{prop} 
 Equation \eqref{eq:singleton} is changed into {\bf connector transitivity}
by replacing the single node $h$ by a connector $\delta$. 
Connector-transitivity extends singleton-transitivity. It concerns  chordless cycles in undirected graphs; the simplest are in Fig. \ref{fig:8}. Furthermore, it concerns   \sUs with mixed edges inducing an undirected $ij$-edge. These have two incoming or two outgoing arrows at $i,j$ and  an 
undirected path via the  nodes of $\delta$. It is the exchangeability property of partial closure which permits one to  argue by just using subgraphs.

Next, we describe  the operator which corresponds closely to partial closure since it  transforms parameter matrices for  Gaussian distributions generated over \Greg in a similar way as partial closure modifies edge matrices.

  \subsubsection*{Partial  inversion}

Let   $N=(1, \ldots, d)$ denote the rows and columns of a real-valued matrix ${\bm M}$
having invertible leading principal submatrices, where $\bm M$ connects 
 real-valued vectors $\bm x$ and $\bm y$ as $\bm{Mx=y}$. The `{\bf partial inversion}' operator, denoted by $\inv_a {\bm M}$, for `$a$' any subset of $N$ and an ordering as $N=(a,b)$,
  exchanges   argument and image
 relating to $a$, \cite{bibWerWieCox06}, \cite{bibWieWer10}, that is
\begin{equation}{\bm M} \left (\begin{array}{r} {{\bm x}_a} \\{\bm x}_b \end{array} \right)=
\left (\begin{array}{r} {{\bm y}_a} \\{\bm y}_b \end{array} \right)
\vspace*{-1mm} \text{ is turned into: \n}
 \inv_a  {\bm M}\left (\begin{array}{r} {\bm{y}_a} \\{\bm x}_b \end{array} \right)=\left (\begin{array}{r} {\bm x}_b  \\{\bm y}_b\end{array} \right)\!. \label{eq:conceptPI}\end{equation} 
 Applied  for instance to rows $a$ of ${\bm \Sigma}^{-1} {\bm X}\!=\!{\bm \zeta}$, two correlated sets of equations turn directly  into  two sets of orthogonalized equations;  see equation \eqref{eq:orthjointr}.\\[3mm]
\noindent{{\bf Definition 5}  The partial inversion operator is, with $a=\{1\}$,
$$ {\bm M}=\begin{pmatrix}  s & {\bm v}\T\\ \bm{w}& \bm m\end{pmatrix}: \nn  \n \inv_{\{1\}}\, \bm{M}=\begin{pmatrix} 1/s  & -\bm{v}\T/s\\ {\bm w}/s& \bm{m}-\bm{wv}\T/s \end{pmatrix}, \label{pinvd}$$ 
and  $\inv_a  {\bm M}$ for $t>1$ elements in $a$,  may be thought of as applying the above  operation $t$ times, by using repeatedly appropriate permutations of $\bm M$.\\[2mm]
 Partial inversion, \cite{bibWerWieCox06}, generalizes the sweep operator, \cite{bibDemp69}, \cite{bibDemp72}, and other methods for  Gaussian elimination, \cite{bibGrcar11}, to non-symmetric matrices.  The matrix $\bm{m}-\bm{wv}\T/s$
 is a Schur complement, \cite{bibSchur1917}.
 A small modi\-fication of the sweep operator leads to  the  `{\bf symmetric difference}', so that 
 an action on $a$, say, is undone by using this same operator again on $a$.

\begin{prop}\label{prop:9}{\em Wermuth, Wiedenbeck and Cox, \cite{bibWerWieCox06}}.
Partial inversion is commutative, can be undone and is exchangeable with taking 
submatrices.\end{prop}

 In particular, 
the operator gives $\inv_b {\bm \Sigma}=-\inv_a {\bm \Sigma}^{-1}$ and  the 
corresponding three Gaussian parameter matrices in equation \eqref{eq:linp}. The Schur complements
involved in the two operations,  $\inv_b {\bm \Sigma}$ and $\inv_a {\bm \Sigma}^{-1}$, are matrix forms 
of the recursion relations in equation \eqref{eq:rconcov}.
A matrix form of the recursion relation for regression coefficients arises by partial inversion on $v$
 in  the matrix example  to equation \eqref{eq:regGpar}. 

By starting from a general  regression graph model in equation \eqref{eq:regGpar},
the parameter matrices   ${\bm H}_{NN}$ and  ${\bm W}_{NN}$ and $N=(u,v)$  are given.
Parameter transformations that are analogous to those of the edge matrices of Prop. \ref{prop:7} 
have been derived using the partial inversion operator and sums of matrix products, \cite{bibWer12}.
In contrast to partial closure, partial inversion may lead to 
negative elements in the induced matrices and therefore permit path cancellations.

\subsubsection*{Some special aspects}

For many regression graph models, the parameters in the regressions of ${\bm X}_k$ given the past  ${\bm X}_{>k}$
 can be estimated by using standard methods, \cite{bibMcCNel89}, \cite{bibWeisb14}, \cite{bibAndSkov10}, but some are based on special
multivariate models, \cite{bibGlonMcCul95}, \cite{bibMarLup11}, \cite{bibRovLupLaR13}, \cite{bibEvansForc14} or  on  special features of the data, \cite{bibEichDahlSan03}, \cite{bibFriedDid03}, \cite{bibCasRov06}.
Possible shortcomings have been identified  for some estimation methods,  \cite{bibMcCull08}, and  for some models, \cite{bibRobinsetal03}, \cite{bibNemRud13}. 
New estimation results are needed for  joint responses of both categorical and quantitative components;  exceptions are CG-regressions, \cite{bibLauWer89},
\cite{bibEdwLau01}.

Features of special models may give unexpected insights  and  often lead to simplified properties. For instance,  parent graphs 
without any transition \sV\!, shown in Fig. \ref{fig:Vdag}, are  lattice conditional independence models, \cite{bibAMPetal97}. 
For these models, \Greg coincides with the ancestor graph. Hence,  the separating  paths of Prop. \ref{prop:4} 
apply directly to \Greg\!.  Parent graphs of exclusively source \sVs are labelled trees, \cite{bibCasSie03}.
These have exactly one path   connecting  each node pair and  $\alpha \ci \beta| c$ if every path between $ \alpha$ and $\beta$
intersects $c$.

Parent graphs without any sink \sVs are said to be decomposable. By  Prop. \ref{prop:5},
they are  Markov equivalent  to  concentration graphs in the same node and edge set.  
Finding well-fitting models for them may often be based on small subsets of variables and, for judging their goodness of fit,  re 
estimation of parameters may not be needed, \cite{bibWer76b}, \cite{bibSundb75}. 
Complex properties of estimates, simplify for decomposable models as well, \cite{bibDawLau93}, \cite{bibLetMas07}. 
Strong analogies to Gaussian models result for binary variables with special types of graph, \cite{bibWerMar14},  
especially when their distributions are  jointly symmetric,  \cite{bibWerMarCox09}, \cite{bibWerMarZwier14}.

For observational studies, it is of concern whether dependences can be well estimated when some variables are unobserved.
As a first step, one  needs to know, when the parameters of such models can be identified. Considerable progress has been made regarding this in the last years; see \cite{bibShpiTian10}, \cite{bibFoyDraiDrton12}, \cite{bibStangVan13},  \cite{bibAllmanetal15}.

\subsubsection*{Some regression graph models for symmetric binary variables}

We now consider special models for symmetric binary variables, which compare most closely to  
Gaussian distributions generated over some regression  graph for variables standardized to have mean zero and unit variance. The purpose is to illustrate for some binary distributions generated over simple Markov equivalent graphs  that the corres\-ponding models are also `{\bf parameter equivalent}',
that is there is a one-to-one relation between the parameters of  two different models. This assures  that the same transformation, which relates  the parameters of the models, applies also to the maximum-likelihood estimates, \cite{bibFisher22}, an important property that appears  not  to be
shared by any of the more  recently  developed estimation methods.

The binary variables have levels ${-1,1}$ and equal probabilities, $\fracshalf$, allocated to 
each of its two levels. A consequence is that they have  mean zero and unit variance by definition.
Their covariance matrix $\bm \Sigma$ coincides therefore with their correlation matrix; it has elements 
$\sigma_{ij}=\rho_{ij}$ and $\sigma_{ii}=1.$ 

Induced marginal and partial correlations 
are just as for Gaussian distributions generated over the same graph, but a zero partial correlation
in an induced concentration graph need not correspond to an independence statement;
for an example see \cite{bibWerMarCox09} Appendix C, and see also \cite{bibLohWain13}.

For an ordered node set $N=(1,2,3,4)$, we denote four  symmetric binary variables by $A,B,C,D$, their respective levels by $i,j,k,l$, and abbreviate joint and conditional probabilities for instance as
 $$ f_{1234}= \pi_{ijkl}^{ABCD}= \Pr(A=i, B=j, C=k, D=l), \nn \n  \pi_{ijk|l}^{ABC|D}=\pi_{ijkl}/\txt \sum_l \pi_{ijkl}.  $$

Their joint distributions are  generated over  parent graphs with just  main effects, for the complete parent graph as:
\begin{eqnarray}  \label{eq:slintri}
\pi_{i|jkl}^{A|BCD}&=& \fracshalf(1+\eta_{12}ij+\eta_{13}ik+\eta_{14} il) \nonumber\\
\pi_{j|kl}^{B|CD}&=& \fracshalf(1+\eta_{23}jk+\eta_{24}jl )\\
\pi_{k|l}^{C| D}&=& \fracshalf(1+\eta_{34}kl )\nonumber\\
\pi_{l}^{D}&=& \fracshalf \nonumber \,, \end{eqnarray}
with
$\eta$'s resulting from $\bm \Sigma$ just  like  linear least-squares regression coefficients:
\begin{eqnarray*} \eta_{34}&=& \rho_{34}\nonumber\\
(\eta_{23}\nn \eta_{24})&=& (\rho_{23}\nn \rho_{24}) {\bm \Sigma}_{\{>2\}\{>2\}}^{-1}\\
(\eta_{12}\nn \eta_{13}\nn \eta_{14})&=& (\rho_{12}\nn \rho_{13}\nn \rho_{14})  {\bm \Sigma}_{\{>1\}\{>1\}}^{-1} \, ,
\nonumber  
\end{eqnarray*}
where the inverse of, say, a submatrix  ${\bm M}_{aa}$ is written as ${\bm M}_{aa}^{-1}.$

This form of the $\eta$-parameters   generalizes directly to $d>4$ variables and stems from the close connection
for  binary variables between probabilities and expectations. For instance, by using equation \eqref{eq:slintri}
$$\E(B|C=k, D=l)=\eta_{23}k+\eta_{24}l,$$
and correlation coefficients are cross-sum differences in probabilities, \cite{bibWerMarZwier14}, such as: 
\begin{equation*} \E(CD)=(\pi^{CD}_{11} +\pi^{CD}_{-1-1})-(\pi^{CD}_{-11} +\pi^{CD}_{1-1})=2(\pi^{CD}_{11}-\pi^{CD}_{-11} )= \rho_{34}. \label{margcor} \end{equation*}
The second equality holds since in the generated  distributions all odd order moments vanish so that 
there is also {\bf joint symmetry}, \cite{bibEdwards00}, Appendix C. In this case, the probability of any level combination  of these binary variables equals the probability of the level combination having each sign switched.

For binary  variables in general,  logit regressions, \cite{bibAndSkov10}, are  best suited to model conditional independence constraints.  A logit regression is   already  close to 
a  linear regression whenever the extreme events are not rare, but instead are  more probable than say 0.1,  \cite{bibCox66}.
It is in the special case of symmetric binary variables that the vanishing of linear regression coefficients in 
equation \eqref{eq:slintri}   coincides with the vanishing of  logit regression coefficients in  corresponding sequences of main-effect logit regressions. Thus, for instance,
$$ 1\ci 4|\{2,3\}\Leftrightarrow (\eta_{14}=0), \fourl  2\ci 3|4\Leftrightarrow (\eta_{23}=0), \fourl  
3\ci 4\Leftrightarrow (\eta_{34}=0)\,.$$

These binary distributions, generated over a given \Gpar, have the edge matrix ${\bm A}$ of equation \eqref{eq:acal}, the same triangular decompositions of ${\bm \Sigma}^{-1}$ and ${\bm \Sigma}$ as  in equation \eqref{eq:icovcon}, and the same induced covariance and concentration graphs as in equation
\eqref{eq:indund}, even though $\bm \Delta$ does not contain the conditional variances but, for $d>2$, their expected values  with respect to the past variables.
We now turn to some 
Markov-equivalent regression graphs and models.\\

{\bf Example 3} The following graph  captures
{\bf mutual conditional independence} of $A,B,C$ given $D$ for $(1,2,3,4)=(A,B,C,D)$ 
\begin{center}
\includegraphics[scale=.58]{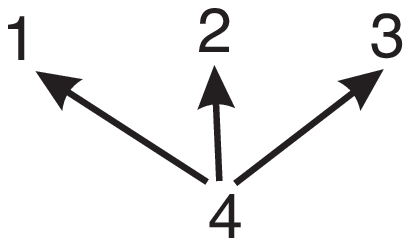}
\end{center}
For any type of distribution generated over this graph,  the  edge matrix $\bm \acal$ is the binary matrix defined by equation \eqref{eq:acal} and the
generated density is 
$$ f_{1234}=f_{1|4} f_{2|4}f_{3|4} f_4  \n  \Leftrightarrow \n (1\ci 2\ci3)|4\,.$$

Here, the four binary symmetric variables have the  constraints $0=\eta_{12}=\eta_{13}=\eta_{23}$ in equations
\eqref{eq:slintri}. The triangular decomposition of $\bm \Sigma$,  that is the matrix pair $(\bm{ A}^{-1}, {\bm \Delta})$, leads to 
the special form of the correlation matrix with
$$ 
\fourl {\bm A}=\begin{pmatrix} 1 &0 & 0  & -\rho_{14}\\
  &1   & 0 &-  \rho_{24}\\  
  & & 1 & - \rho_{34}\\  {\bf 0}&&&  1
\end{pmatrix}, \fourl
 {\bf \Sigma}=\begin{pmatrix}   1   & \rho_{14} \rho_{24}&\rho_{14}\,\rho_{34}& \rho_{14}\\  
   . & 1 &\rho_{24}\rho_{34}& \rho_{24}\\  .&.&1& \rho_{34}\\  .&.& .& 1
\end{pmatrix} $$
 and  $ \delta_{ss}=1-\rho_{s4}^2$ for $s=1,2, 3$, and $\delta_{44}=1 $. The induced 
 correlations, corresponding to the three missing edges of the graph, are as specified for 
the outer nodes of a  source \sV in equation \eqref{eq:slintri}. Here every ancestor is a parent,  hence $\acal^{-} =\acal$,
and equation \eqref{eq:indund}  gives a complete
  induced covariance graph and an induced concentration graph
 with no additional edge.

Because the given \Gpar
 contains no collision \sV\!, it is Markov equivalent to  \Gcon with the 
 same node and edge set. The joint probabilities obtained from equations \eqref{eq:slintri}
 show directly that  the more important  parameter equivalence holds in addition.  
Often, Markov equivalence  implies  parameter equivalence whenever 
a single parameter is attached to each edge present in \Gpar\!.\\

\noindent {\bf Example 4} This  example  is a {\bf Markov chain}
graph, which is a  parent graph consisting of a single direction-preserving  path of  arrows, here: 
$$ 1\fla 2 \fla 3 \fla 4\, ,$$
where each response node remembers from its past only the most recent node. 
For any type of distribution generated over this graph,  the  edge matrix $\bm \acal$ is the binary matrix defined by equation \eqref{eq:acal} and the
generated density is 
$$ f_{1234}=f_{1|2} f_{2|3}f_{3|4} f_4  \n  \Leftrightarrow \n (1\ci \{3,4\}|2 \text{ and } 2\ci 4|3)\,.$$

For four binary symmetric variables and constraints  $0=\eta_{13}=\eta_{14}=\eta_{24}$ in equations
\eqref{eq:slintri}, the triangular decomposition of $\bm \Sigma$,  the matrix pair $(\bm{ A}^{-1}, {\bm \Delta})$ gives the special form of the correlation matrix with
$$  {\bm A}=\begin{pmatrix} 1 &-\rho_{12} & 0  & 0\\
  &1   &- \rho_{23} & 0\\  
  &  & 1 &-\rho_{34}\\  {\bf 0}&&&  1
\end{pmatrix}, \nn \n {\bm \Sigma}=\begin{pmatrix} 1 &\rho_{12} &\rho_{12}\,\rho_{23}  &\rho_{12}\,\rho_{23}\,\rho_{34}\\
  . &1   & \rho_{23} &\rho_{23}\,\rho_{34}\\  
  . & . & 1 &\rho_{34}\\ .&.&.& 1
\end{pmatrix}\, ,$$
and  $ \delta_{ss}=1-\rho_{s, s+1}^2$ for $s=1,2, 3$ and $\delta_{44}=1 \, .$ The correlation induced 
for each missing edge in \Gpar equals the product of the correlations along the path connecting the uncoupled node pair.  Here, every node in the past of $i$ is an ancestor of $i$, hence the ancestor graph with edge matrix ${\bm \acal}^{-}$ is complete. Consequently, the induced
covariance graph is also complete.

As in Example 3, equation \eqref{eq:indund}  gives an induced concentration graph
 with no additional edge. Here,  \Gpar is Markov equivalent to  a \Gcon which 
is a  {\bf concentration chain} in nodes $(1,2,3,4)$: 
 $$ 1\ful 2 \ful  3 \ful 4, $$ 
 where  each edge present means $i\pitchfork j|N\setminus \{i,j\}$. Also as in Example 1,
 there is parameter equivalence obtained from equation \eqref{eq:slintri} to the parameters
 in the joint distribution:
$$ \pi_{ijkl}^{ABCD}=\fracssixteenth\{(1+\rho_{12}ij) (1+\rho_{23}jk) (1+\rho_{34}kl) \, .$$ \n\\[-6mm]

\noindent{\bf Example 5} This  last example is quite different from the previous one. It is  a {\bf covariance chain} in nodes $(1,2,3,4)$:
$$ 1\dal 2\dal 3\dal 4\, , $$
where each $ij$-edge present represents in general $i\pitchfork j$. For the symmetric binary variables, the dependence is captured by  the marginal correlation coefficient, $\rho_{ij}\neq 0$.
 The simplifying 
independences are in $\bm A^{-1}$ and in $\bm \Sigma$ but there are none for the joint distribution
generated over a parent graph with node ordering $(1,2,3,4)$.
Accordingly, the factorizations and the independence structure are
$$ (f_{124}=f_{12} f_4 \text{ and } f_{134}=f_1f_{34}) \Leftrightarrow (\{1,2\} \ci 4 \text{ and } 1\ci \{3, 4\})$$

Because dashed-line \sVs are edge-inducing by conditioning, induced regression coefficients appear in $\bm A$, where   ${\bm A}\T{\bm \Delta}^{-1}{\bm A}={\bm \Sigma}^{-1}$,  
\begin{equation}
 {\bm A}= \begin{pmatrix} 1 & -\eta_{12} & \eta_{12} \,\eta_{23}& -\eta_{12}\,\eta_{23}\,\eta_{34}\\
                                           & 1&-\eta_{23}& \eta_{23}\,\eta_{34}\\
                                           &    & 1& -\eta_{34}\\
                                            {\bm 0}& & &1 \end{pmatrix} \nn
{\bm \Sigma}= \begin{pmatrix} 1 &\rho_{12} & 0  & 0\\
  . &1   & \rho_{23} & 0\\  
  . & . & 1 &\rho_{34}\\ .&.&.& 1, \end{pmatrix} \nn \,.\label{eq:ex5}\end{equation}
  Since there are no vanishing regression coefficients, there are also
  no independences of the type $i\ci j |N\setminus\{i,j\}$: hence the induced concentration graph is complete.

   There can be sign changes for induced coefficients, for instance $a_{24}=- a_{23}a_{34}$.
    Separate estimation of the parameters in the
   regressions for responses $i<3$ 
   is not feasible for this model, since some of the regression coefficients depend on coefficients in the past of node $i$.                               
                                         
However by Prop. \ref{prop:5}, the given covariance chain is  Markov equivalent to the following regression graph
$$ 1\fra 2\dal 3\fla 4\, , $$
which represents, 
for Gaussian distributions,  the simplest type of Zellner's, \cite{bibZellner62}, \cite{bibDrton09}, seemingly unrelated regression. After reordering to $(2, 3, 1,  4)$, the covariance matrix here becomes  ${\bm \Sigma} '$
while partial inversion on the regressors $1,4$ gives the parameters for the joint response regression of ${\bm Y}_a$ on ${\bm Y}_b$, where $a=\{2,3\}$ on $b=\{1,4\}$
$$ {\bm \Sigma}'=\begin{pmatrix}\! 1 &\rho_{23} &    \rho_{12} & 0\\
  . &1   &0& \rho_{34}  \\  
  . & . & 1 &0 \\ .&.&.& 1 
\end{pmatrix}\nn  \inv_{3,4}\,{\bm \Sigma}'=\begin{pmatrix}\! 1 -\rho_{12}^2&\rho_{23} &  \rho_{12}& 0\\
  .&1-\rho_{34}^2   & 0&\rho_{34} \\  
  \sim & \sim& 1 &0 \\ \sim& \sim&.& 1\end{pmatrix} .
$$ The $\sim$ notation denotes entries that are symmetric up to the sign.   
This parametrization is equivalent to requesting
$$ 2\pitchfork 3|\{1,4\}, \nn 2\pitchfork 1|4, \nn 2 \ci4|1, \nn  3\ci 1|4, \nn 3\pitchfork  4|1, \nn 1\ci 4.$$
It leads to joint probabilities which become, see \cite{bibWerMarCox09} Appendix A:
$$\pi_{jkil}^{BC AD}=\fracssixteenth(1+\rho_{12}ij+\rho_{23}jk+\rho_{34}kl +\rho_{12}\rho_{34} ijkl) \,,$$
so that there is  parameter equivalence in spite of a   four-factor interaction.

For $N=(2,3,1,4)$, the  three types of edge sets, $ E_{\dals},E_{\flas},  E_{\fuls} $ are captured by 
$${\bm \hcal} =\begin{pmatrix} \,1& 0& 1 & 0 \\ \, 0&  1& \,0  & 1 \\
\,0&0&1&0\\
\,0&0&0&1 \end{pmatrix}, \nn
{\bm \wcal}=\begin{pmatrix} \,1 & 1 & 0& 0\,\\ \, 1& 1&0 & 0\\
\,0&0&1&0\\
\,0&0&0&1 \end{pmatrix} . 
  $$
  
  Here, ${\bm \ecal}=\In({\bm \wcal}+{\bm \hcal}+{\bm \hcal}\T)$ is the generating edge matrix of ${\bm \Sigma}$, due to  Markov equivalence. With it, the edge matrix of the induced concentration graph becomes, by equation \eqref{eq:indcovcon},       ${\bm S}^{NN}=\In({\bm L}^{-1})$, where ${\bm L}= 5{\bm \ical}-{\bm \ecal}$. This is just one way to see  that here
the induced concentration graph is complete.

  These last three examples show that independences may mean simple zero constraints
on parameters of one model but may appear  as complex constraints, even in parameter
equivalent models. It is therefore in general rarely useful
to restrict model search and data analysis to one particular class of models. Strong prior knowledge would be an exception.  Even then, using Markov- and parameter-equivalence  may  aid in finding alternative interpretations  and alternative  fitting algorithms.

If the motivation for designing an empirical study are causal hypotheses, then undirected graphical models alone are typically of little interest.
But similarly, directed acyclic graph models are of little help when one expects that an intervention will lead to changes in several connected responses
at the same time.  For instance,  when effects of a drug to reduce blood pressure are to be studied,   this intervention will affect systolic and
diastolic blood pressure  simultaneously and not one  before the other.

\subsubsection*{Discussion}

It took nearly 40 years of research until the present form of the regression graph, \Greg\!, was defined and its properties and consequences could be studied.
The graph represents ordered sequences of joint response regressions. Responses  may depend on all or  on only some of the variables in their past.  The graph contains three types of edge, one undirected type for dependences among responses, another undirected   type
for dependences among context variables and directed edges pointing  to a response from nodes in its past.  Conditional dependences  show in edges present in \Greg.\! These dependences simplify with  more missing
edges, that is the more conditional independence constraints there are. 

To make the graphs useful tools for tracing developmental pathways and for predicting structure in alternative models, the generated distributions have to mimic some properties of joint Gaussian distributions, see Prop.\,\ref{prop:1}.
 If in \Greg\!, independences did not combine downwards and upwards, that is if the intersection 
and the composition properties were not satisfied, it would  be impossible to infer mutual independence of disconnected subgraphs.  Then, the graphical representations would be nearly useless. 

If regression graphs  were not, in addition, singleton-transitive,  then they would not even well represent 
Gaussian distributions which  have this property and are the simplest and  most studied types of joint distribution. Also, edges present in  induced graphs would not  point to  non-vanishing conditional dependences in traceable regressions. But this is a prerequisite for  useful tracings of  pathways  of development in the graphs.
 
Connector-set transitivity will illuminate  the distinction between structural independences  and 
those that may result  due to special  parametric constellations. The distinction between the two types reflects a long-standing 
practice in empirical research. Whenever a result has been replicated in several studies under essentially the same conditions,  one typically still wants to establish
it  under  modified conditions.

Even when all edges present in a graph correspond to positive dependences,
negative linear dependences are induced by closing collision \sVs\!; see equation \eqref{eq:indr} and equation \eqref{eq:ex5}. Some first results
for preserving positive dependences have been obtained, \cite{bibMaXieGeng06}, \cite{bibJiangDingPend15}, others are expected for totally positive distributions generated over \Gcon and for decomposable regression graphs, those without any collision \sV\!.

Among the open theoretical questions are the following: 
Can necessary and sufficient conditions be derived for the properties of traceable regressions, such as those for the intersection property, \cite{bibSanMMouRol05}?  
 For this, can  methods of algebraic statistics also be helpful, such as those for binary tree models, \cite{bibZwierSmith11}? 
 How will independence 
structures and their properties  change when  graphs are no longer finite, \cite{bibMontRaja15}?
When may models with fewer independence constraints, that is with more edges in the graph, be safely used as covering models, \cite{bibCoxWer90},
for simpler estimation and  useful interpretation?

At least equally important are further direct applications of traceable regressions; for a summary of the related tasks and links to detailed reports on
finding well-fitting models in different research contexts see \cite{bibWerCox15}. With traceable regressions, it has become feasible, for the first time, to derive the structural consequences of (1) ignoring some of the variables,  of (2) selecting subpopulations via fixed levels of some other variables or of (3)
changing the order in which the variables might get generated.
With the currently used methods for combining results from empirical studies, called `{\bf meta-analyses}', such effects are not taken care off.  Therefore, the, most important future  applications of these models  will aim at the best possible integration of knowledge from related studies.

 \renewcommand{\baselinestretch}{0.8}
\renewcommand\refname
\end{document}